\documentclass[twocolumn,superscriptaddress,amssymb, nobibnotes, aps, prb]{revtex4-2}
\usepackage{enumitem}
\usepackage{romannum}
\usepackage{amsthm}
\usepackage{hhline}
\usepackage{epsfig,amsopn}
\usepackage{graphicx}
\usepackage{xcolor}
\usepackage[english]{babel}
\usepackage{amsmath,amssymb}
\usepackage[colorlinks,allcolors=blue]{hyperref}
\usepackage{url}
\usepackage{multirow}
\usepackage{enumerate}
\usepackage{verbatim}
\usepackage{bm}
\begin{document}
\title{Topology of multipartite non-Hermitian one-dimensional systems} 
\author{Ritu Nehra}
\author{Dibyendu Roy}
\affiliation{Raman Research Institute, Bangalore-560080, India}
\begin{abstract}
The multipartite non-Hermitian Su-Schrieffer-Heeger model is explored as a prototypical example of one-dimensional systems with several sublattice sites for unveiling intriguing insulating and metallic phases with no Hermitian counterparts. These phases are characterized by composite cyclic loops of multiple complex-energy bands encircling single or multiple exceptional points (EPs) on the parametric space of real and imaginary energy. We show the topology of these composite loops is similar to well-known topological objects like Möbius
strips and Penrose triangles, and can be quantified by a nonadiabatic cyclic geometric phase which includes contributions only from the participating bands. We analytically derive a complete phase diagram with the phase boundaries of the model. We further examine the connection between encircling of multiple EPs by complex-energy bands on parametric space and associated topology.

\end{abstract}
\maketitle
\section{Introduction}
The topological understanding of non-Hermitian systems has recently generated enormous interest in various scientific disciplines including photonics, condensed matter physics and acoustics~\citep{jin_topological_2017,longhi_non-hermitian_2015,ghatak_observation_2020,jiang_multiband_2021,berry_physics_2004,cartarius_exceptional_2009,xu_topological_2016,cartarius_exceptional_2009}. 
The presence of dissipation, e.g., friction, viscous drag, resistance, and pumping of energy or particle to a system make it non-Hermitian. Standard lasers functioning using population inversion and metallic electrons inside thermal phonon environment are some examples of non-Hermitian systems, which are in abundance in nature. Therefore, an extension of the concepts of topology in Hermitian systems to their non-Hermitian counterparts is vital~\citep{wagner_numerical_2017,gong_topological_2018,kawabata_parity-time-symmetric_2018,okugawa_non-hermitian_2021,hatano_localization_1996}. While energy of the bands of Hermitian lattices is real, it is generally complex for non-Hermitian lattices~\citep{hatano_localization_1996,okugawa_non-hermitian_2021,wang_topological_2021,hu_knots_2021,PhysRevLett.124.056802}. The last feature of non-Hermitian systems leads to special type of singularities or branch cut points called exceptional points (EPs)~\citep{heiss_physics_2012,berry_physics_2004,heiss_chirality_2008,guenther_projective_2007,dembowski_experimental_2001}  where the eigenvalues and eigenfunctions coalescence~\citep{zhang_correspondence_2020,heiss_chirality_2008,heiss_repulsion_2000,dembowski_encircling_2004,heiss_phases_1999}. 
The encircling  of an EP on the parametric space and associated topology has been studied and observed in micro-disks  experiments~\citep{ryu_analysis_2012,doppler_dynamically_2016,hassan_dynamically_2017,lee_anomalous_2016}.

Previous works have mostly focused on topological properties, such as topological phases and invariants to characterize those phases, of non-Hermitian bipartite systems~\citep{lieu_topological_2018,vyas_topological_2021,yin_geometrical_2018,shen_topological_2018,fu_extended_2020}.  It has been suggested that topological phases of non-Hermitian systems should be understood as {\it dynamical phases} by taking into consideration the full complex spectra of the systems rather than individual eigenstates \citep{gong_topological_2018,kawabata_symmetry_2019}. A global (complex) Berry phase has been introduced as topological invariant identifying different topological phases of the non-Hermitian systems~\citep{liang_topological_2013}. The braiding of two non-Hermitian bands is measured by a global topological invariant (related to vorticity formula), which counts how many times two bands braid in the complex energy-momentum space as momentum $k$ varies from 0 to $2\pi$~\citep{wang_topological_2021}. There are also many recent studies exploring topological features of multiband non-Hermitian one-dimensional (1D) systems~\citep{lieu_topological_2018-1,jin_topological_2017,liu_new_2017,martinez_alvarez_edge_2019,he_topology_2018,ryu_topological_2002,zhu_mathcalpt_2014,lieu_topological_2018,vyas_topological_2021,yin_geometrical_2018}. Yet, it is still not clear from these studies : (1) How to quantify topology of gapless and gapped phases in multipartite non-Hermitian 1D systems?  and (2) Is there a unique connection between encircling of multiple EPs in complex-energy parametric space and associated topology?



To find the answer to the above questions, we here investigate a multipartite generalization of the non-Hermitian Su-Schrieffer-Heeger (SSH) model~\citep{su_soliton_1980,zhu_floquet_2020,pan_photonic_2018,bergholtz_exceptional_2021,li_topological_2020,zhang_exceptional-point-induced_2019,jin_topological_2017,parto_edge-mode_2018,jiang_multiband_2021,fu_extended_2020,han_topological_2021,wu_topology_2021} as a prototypical example. Our generalized model falls in the equivalent topological classes AI or D$^\dagger$ with $\mathcal{S}_+$ sublattice symmetry of the non-Hermitian symmetry classes~\citep{kawabata_symmetry_2019}. Contrary to the previous assertion \citep{lieu_topological_2018-1}, we notice that the value of complex Berry phase for individual bands is not quantized for multipartite non-Hermitian SSH model even in the presence of sublattice symmetry. Therefore, it can not act as a good topological invariant to classify topological phases in such systems. We further observe that the global Berry phase is also inadequate to classify topological phases and phase transitions in multipartite systems since it includes contributions from bands that may not participate in the particular phases and transitions~\citep{liang_topological_2013}. For similar reasons, a global vorticity formula fails to characterize the topological phases~\citep{wang_topological_2021,shen_topological_2018,ghatak_new_2019,ghatak_observation_2020,yin_geometrical_2018}. Thus, a systematic description is required to construct useful topological invariants in multipartite non-Hermitian systems.

We here revisit the complex energy spectra ($E(k)$) of the multipartite non-Hermitian model on parametric space of real and imaginary $E(k)$ \citep{gong_topological_2018}, where they form loops that can encircle single or multiple EPs. For some parameter regimes, each band forms a separate loop that does not encompass any EP (see Figs.~\ref{fig:intro}(a,e)). These separate loops represent the topologically trivial and nontrivial gapped insulating phases similar to those in the corresponding Hermitian SSH model~\citep{su_soliton_1980,asboth_su-schrieffer-heeger_2016,nehra_transmission_2021} and we show below that these phases can still be characterized by complex Zak phase~\citep{vyas_topological_2021,lieu_topological_2018,liang_topological_2013} of individual bands. When loops in Fig.~\ref{fig:intro} are formed by two or multiple bands, they enclose single or multiple EPs. The encircling of such EPs on the parametric space (Re[$E(k)$], Im[$E(k)$]) leads to an exchange of eigenmodes indicating breakdown of adiabatic cyclic evolution of a state over the first Brillouin zone. Thus, the complex Zak phase for individual bands is ill-defined and not quantized for such composite loops. These loops represent composite insulating and metallic phases with no Hermitian counterparts. As we explain below, we can still define a geometric phase for a cyclic nonadiabatic quantum evolution along each composite loop formed by multiple bands. 
The value of the geometric phase for the composite loops can be quantized and non-zero for a topologically nontrivial composite phase. Interestingly, we find that the value of the geometric phase is independent of the number of EPs encircled by the composite loops specifying the contour for the geometric phase. 

The rest of the paper is divided into five sections. In Sec.~\ref{model}, we introduce the multipartite non-Hermitian SSH model and give the complex energy spectra of its quadripartite version. In Sec.~\ref{para}, we show interesting features of complex energy spectra of the quadripartite non-Hermitian SSH  model on parametric space of real and imaginary energy. We discuss the phase boundaries of the quadripartite non-Hermitian SSH  model in Sec.~\ref{pb} and propose a bi-orthonormal geometric phase for the composite phases in Sec.~\ref{geo}. We conclude in Sec.~\ref{con}.  We further add six appendices for (i) topological details of the bipartite and tripartite non-Hermitian SSH model, (ii) analytical details of the hybrid point (HP) and EP calculations, (iii) dispersion spectra and the other topological phases of the quadripartite non-Hermitian SSH  model, and (iv) other non-Hermitian properties like symmetries, vorticity and braiding in all three models.

\section{Model}
\label{model}
Let us consider the following Hamiltonian of the multipartite non-Hermitian SSH model of $L$ unit cells with $s$ sites per unit cell~\citep{vyas_topological_2021,yin_geometrical_2018,shen_topological_2018}:  
\begin{align}
H_{s}=\displaystyle\sum_{j=1}^{L-1}&\Big(\sum_{\sigma=1}^{s-1}(t_{\sigma r}c_{j,\sigma+1}^\dagger c_{j,\sigma}+t_{\sigma l}c^\dagger_{j,\sigma}c_{j,\sigma+1})\nonumber\\&+(t_{sr}c^\dagger_{j+1,1}c_{j,s}+t_{sl}c^\dagger_{j,s}c_{j+1,1})\Big),  \nonumber\\&+\eta(t_{sr}c^\dagger_{1,1}c_{L,s} +t_{sl}c^\dagger_{L,s}c_{1,1}),
\label{eq:1}
\end{align}
where $c^\dagger_{j,\sigma}(c_{j,\sigma})$ denotes a spinless fermionic creation (annihilation) operator on $\sigma^{th}$ sublattice site of $j^{th}$ unit cell of the 1D system. Here,  $t_{\sigma r}$ is the right moving hopping from $\sigma^{th}$ sublattice site to the right-next site and $t_{\sigma l}$ is the left moving hopping from right-next site to the $\sigma^{th}$ sublattice site. These nonreciprocal couplings ($t_{\sigma r}$, $t_{\sigma l}$) are solely responsible for the non-Hermiticity in the above model in contrast to gain/loss models~\citep{wu_topology_2021,zhu_mathcalpt_2014,liu_metrology_2016,midya_non-hermitian_2020}. The impact of boundaries on the physics of non-Hermitian systems  can be tuned in the above Hamiltonian by the parameter $\eta$, which takes a value as $0,1,\frac{1}{L}$ for open, periodic, and special boundary conditions, respectively. One highly explored feature of non-Hermitian systems is the violation of conventional bulk-boundary correspondence due to various skin effects in these systems, such as non-Hermitian skin effect~\citep{okuma_topological_2020,yuce_non-hermitian_2020}, tidal skin effect~\citep{zhang_tidal_2021}, critical skin effect~\citep{li_critical_2020}, hybrid higher-order skin effect~\citep{zou_observation_2021}. 
However, the non-Hermitian Bloch band theory generalizes the Brillouin zone~\citep{imura_generalized_2020}, which restores the bulk-boundary correspondence in non-Hermitian topological systems. We show below a restoration of bulk-boundary correspondence in the multipartite non-Hermitian SSH model using the special boundary conditions \cite{vyas_topological_2021}.

\begin{figure}[h!]
\centering
\includegraphics[width=\linewidth, height=0.75\linewidth]{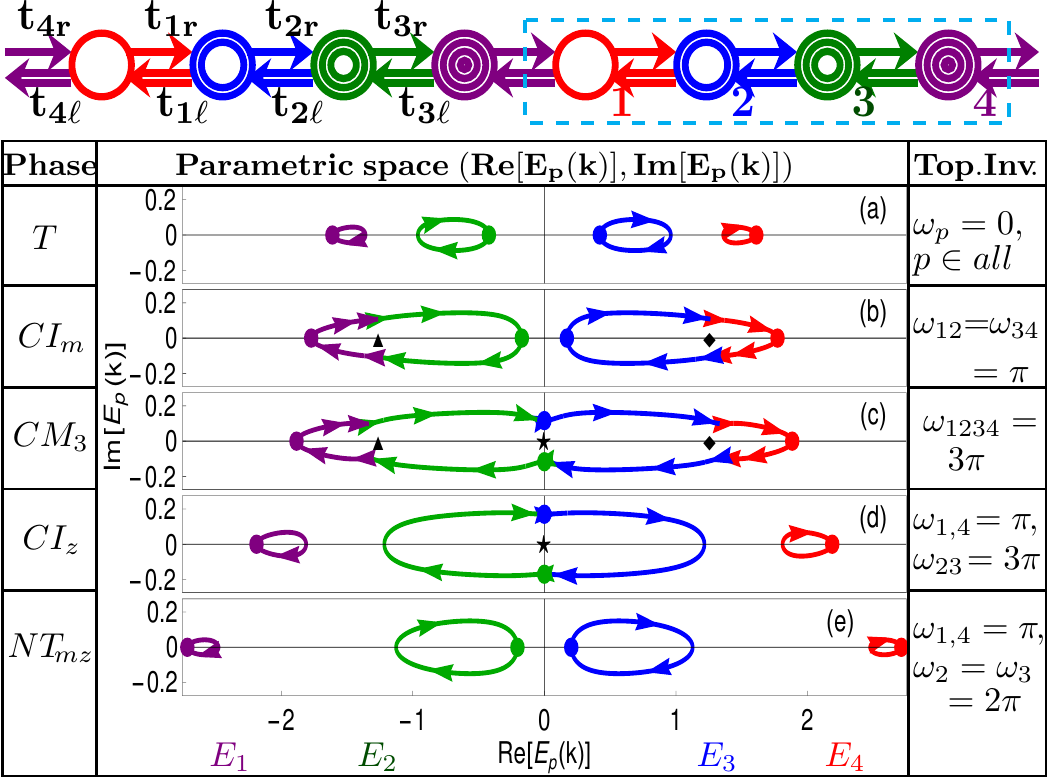}
\caption{(Top) A cartoon of the quadripartite non-Hermitian SSH model showing various hopping amplitudes. (Below) Different types of loops on the parametric space of real and imaginary energy formed by four complex energy spectra $E_p(k)$ of the above model as the inter-cell hopping is tuned. The topological invariants (TIs), $\omega_p, \omega_{p_1p_2\dots p_n}$, characterize various phases represented by these loops encircling exceptional points ($\blacktriangle, \blacklozenge, \bigstar$). The parameters are $t_{1r}=t_{1l}=t_{3r}=t_{3l}=t=1$, $t_{2r}=t_{2l}=t_2=0.8$, $t_{4r}=t_{4l}e^{-\theta}=t_4,\theta=0.75$ and $t_4=0.25,\;0.5,\;0.65,\;1,\;1.5$ in (a,b,c,d,e) respectively.}
\label{fig:intro}
\end{figure}
The bipartite ($s=2$) version of the above Hamiltonian in Eq.~\ref{eq:1} was explored recently in Ref.~\citep{vyas_topological_2021} to reveal appearance of a gapless M{\"o}bius metallic phase due to non-Hermiticity. The nontrivial topological properties of this metallic phase can be characterized by a global bi-orthonormal geometric (Zak) phase by extending the notion of circuit in $k$ space as a loop from $k = 0$
to $k = 4\pi$. We provide further details including parametric energy plot and braiding of complex energy bands of the bipartite model in Appendix~\ref{nssh2}. Later, we briefly discuss appearance of a trivial metallic phase due to non-Hermiticity in tripartite non-Hermitian SSH model and add details of this case in Appendix~\ref{nssh3}.

We here primarily explore $s=4$ case of non-Hermitian SSH model shown in Fig.~\ref{fig:intro}. Employing translational symmetry with periodic boundary condition, we get four complex-energy spectrums $E_{p}(k)$ of the model:
 \begin{align}
E_{p}(k)= \pm\sqrt{\alpha\pm\sqrt{\alpha^2-\beta}},
\label{eq:15}
\end{align}
 where $\beta=(t_{4r}t_{2r}-t_{1l}t_{3l}e^{-ik})(t_{4l}t_{2l}-t_{1r}t_{3r}e^{ik}),\;\alpha=(t_{1l}t_{1r}+t_{2l}t_{2r}+t_{3l}t_{3r}+t_{4l}t_{4r})/{2}$ and the band index $p=1,2,3,4$ are denoted by $-+,--,+-,++$, respectively. For the sake of simplicity in the following discussion, we choose a set of parameters with intrinsic inversion symmetry~\citep{zak_berrys_1989,xiao_coexistence_2017} (i.e., inversion symmetry within the unit cell) as $t_{1l}=t_{1r}=t_{3l}=t_{3r}=t$ , $t_{2r}=t_{2l}=t_2$ and $t_{4l}e^{-\theta}=t_{4r}=t_{4}$, where $\theta$ controls non-Hermiticity of the model \footnote{The intrinsic inversion symmetry establishes an inversion axis at the center of the quadripartite unit cell (midpoint between sub-lattice site $\sigma=2$ and $3$) in such a way that the hoppings are equal to the corresponding mirror reflection hoppings which leads to the relations $t_{1l}=t_{3r}$, $t_{1r}=t_{3l}$, $t_{2l}=t_{2r}$.}. The complex energy dispersion spectra for such parameters are shown in Appendix~\ref{cDispersion}. 
 The Hermitian quadripartite SSH model~\citep{xie_topological_2019,maffei_topological_2018,eliashvili_edge_2017} ($\theta=0$) has two topological phase transitions with increasing $t_4$ (when $t_1t_3>t_2^2$): (a) first at $t_4=t_2$ and $t_1=t_3$ when topology of all four bands switches from trivial to nontrivial, and (b) second at $t_4=t_1t_3/t_2$ when middle two bands return to topologically trivial phase (Appendix~\ref{cHermitian}). These topological phases and phase transitions in the Hermitian model can be characterized by real Zak phase~\citep{berry_adiabatic_1987,atala_direct_2013,asboth_su-schrieffer-heeger_2016} which we show by dashed lines in Fig.~\ref{fig:phase1}. 

\section{Parametric energy plots}
\label{para}
The parametric energy plots of the four complex-energy bands over first Brillouin zone $k\in[0,2\pi)$ are shown in Fig.~\ref{fig:intro} for different inter-cell hopping $t_4$ and fixed intra-cell hoppings $t,t_2$ as well as non-Hermiticity $\theta$. The bands with real-energy gaps from each other create separate loops on the parametric space which are shown in Figs.~\ref{fig:intro}(a,e) for a small and large value of $t_4$. These separate loops are reminiscent of topologically trivial ($T$) and nontrivial ($NT$) gapped insulating phases in the Hermitian counterpart. We find that the topological properties of these bands can be described by complex Zak phase for a cyclic adiabatic evolution over each separate loops ~\citep{gong_topological_2018,ghatak_new_2019,zhang_partial_2019,midya_non-hermitian_2020}:
\begin{align}
\omega_{p}=i\displaystyle\int_{0}^{2\pi}\phi_p(k).\partial_k\psi_p(k)\; dk,
\label{eq:5}
\end{align}
where $p$ represents the band index, and $\phi_p(k)$ and $\psi_p(k)$ are left and right eigenvectors of the Hamiltonian $H_4$ in the momentum space. 
The gauge-invariant form of the complex Zak phase is defined as $\tilde{\omega}_p=\omega_p(\text{mod}\;2\pi)$ due to single-valuedness of $\phi_p(k)$ and $\psi_p(k)$. The set of bi-orthonormal eigenvectors associated with complex eigenenergy bands ($p=1,2,3,4$) in Eq.~\ref{eq:15} of the non-Hermitian SSH model reads as 
\begin{align}
&\phi_p(k)\hspace{-2pt}=\hspace{-2pt}\frac{1}{\sqrt{N_p}}
\begin{bmatrix}
E_p(k)^2t_{4l}e^{-ik}+t_{2r}(t_{1r}t_{3r}-t_{2l}t_{4l}e^{-ik}) \\
E_p(k)(t_{1l}t_{4l} e^{-ik}+t_{2r}t_{3r}) \\E_p(k)^2t_{3r}-t_{1l}(t_{1r}t_{3r}-t_{2l}t_{4l}e^{-ik})
\\E_p(k)(E_p(k)^2-t_{1r}t_{1l}-t_{2r}t_{2l}) 
\end{bmatrix}^T\hspace{-10pt}, \nonumber \\
&\psi_p(k)\hspace{-3pt}=\hspace{-3pt}\frac{1}{\sqrt{N_p}}\hspace{-3pt}
\begin{bmatrix}\hspace{-1pt}
E_p(k)^2t_{4r}e^{ik}+t_{2l}(t_{1l}t_{3l}-t_{2r}t_{4r}e^{ik}) \hspace{-1pt}\\
E_p(k)(t_{1r}t_{4r} e^{ik}+t_{2l}t_{3l}) \\E_p(k)^2t_{3l}-t_{1r}(t_{1l}t_{3l}-t_{2r}t_{4r}e^{ik})\\
E_p(k)(E_p(k)^2-t_{1r}t_{1l}-t_{2r}t_{2l}) 
\end{bmatrix}\hspace{-3pt},
\label{eq:19}
\end{align}
with a normalisation constant $N_p(k)=2E_p(k)^2(E_p(k)^2-t_{1r}t_{1l}-t_{2r}t_{2l})(2E_p(k)^2-(t_{1r}t_{1l}+t_{2r}t_{2l}+t_{3r}t_{3l}+t_{4r}t_{4l}))
$. The values of $\omega_p$ are always real for our sublattice symmetric SSH model~\citep{lieu_topological_2018}.  We find $\omega_p=0$ for all loops in Fig.~\ref{fig:intro}(a) indicating topologically trivial gapped insulating phase for all four bands at this parameter regime. For the loops in Fig.~\ref{fig:intro}(e), we find $\omega_{1,4}=\pi$ and $\omega_{2,3}=2\pi~(\tilde{\omega}_{2,3}=0)$ which tell while the middle two bands ($p=2,\;3$) are in topologically trivial gapped insulating phase, the other two bands ($p=1,\;4$) are in topologically nontrivial gapped insulating phase.  The values of $\omega_p$ for individual bands in different topological phases are given in Fig.~\ref{fig:phase1}.

The energy bandgap vanishes for some specific choice of parameters which can be calculated by finding the energy band touching conditions.  From Eq.~\ref{eq:15}, we observe that the bands of non-Hermitian SSH model touch each other when (1) $\alpha^2-\beta=0$, leading to touching of  ($E_1\;,E_2$) and ($E_3\;,E_4$) simultaneously at $k=\pi$, and (2) $\beta=0$ resulting in $E_2$ touching $E_3$ at $k=0,2\pi$.  These band touching occurs at different values of $k$(=0, $\pi$, $2\pi$) and $E_p(k)$, which cause three second-order EPs created by the band touching of any two bands (Appendix~\ref{cEP}). But, for some other choice of parameters, we can have all four bands touch at a same energy ($E_p(k)=0$) and momentum $k=\pi$, leading to a single fourth-order EP (Appendix~\ref{cfourth}).


In Fig.~\ref{fig:intro}(b), the two upper ($E_3,E_4$) and two lower ($E_1,E_2$) bands separately touch at constant real energy values for $k=\pi$ with a discontinuity in imaginary energies at the same points which are EPs ($\blacktriangle, \blacklozenge$). Likewise, the middle pair of energy bands $(E_2,E_3)$ form a composite loop and enclose the third EP ($\bigstar$) at zero energy in Fig.~\ref{fig:intro}(d) for another value of $t_4$. These composite loops of multiple (real-energy) gapless bands in Figs.~\ref{fig:intro}(b) and (d) represent insulating state due to the presence of real-energy gap(s) with the other bands, and we denote these composite insulating states by $CI_m$ and
$CI_z$ (where the subscripts $m$ and $z$ stand for mid-gap energy
and zero energy states), respectively. Interestingly, all three second-order EPs ($\blacktriangle, \blacklozenge, \bigstar$) are enclosed simultaneously at some intermediate $t_4$ when all four bands together generate a cyclic contour in Fig.~\ref{fig:intro}(c). This composite loop in  Fig.~\ref{fig:intro}(c) depicts a composite metallic phase encircling  three EPs assigned as $CM_3$. The band-structure topology of the contributing bands in the composite insulating and metallic states is similar to the M{\"o}bius strips and the Penrose triangles, respectively. In Appendix~\ref{spenrose}, we explain how to realize topology of two-band and four-band composite loops with $4\pi$ and $8\pi$ periodicity on the parametric space using M{\"o}bius strips and  Penrose triangles, respectively. Apart from real-energy band touching, the eigenvector~\citep{dembowski_encircling_2004,heiss_phases_1999} and the imaginary energy of the bands are also exchanged while encircling EPs by composite loops. 
The encircling of one and three EPs by composite loops in Figs.~\ref{fig:intro}(b,d) and Fig.~\ref{fig:intro}(c) lead to an increase in the periodicity of a quantum-state evolution to $4\pi$ and $8\pi$, respectively~\citep{ryu_analysis_2012,heiss_chirality_2008}. Similar enhancement of periodicity also occurs for other composite loops in bipartite ($s=2$) and tripartite ($s=3$) non-Hermitian model depending on number of encircled EPs (Appendices~\ref{nssh2},~\ref{nssh3}). Therefore, to describe the topological properties of such composite insulating phases represented by composite loops, we must consider only those participating bands in the composite loops instead of all bands as in global complex Berry/Zak phase. 
The braiding of these complex-energy bands can be understood by plotting them in three-dimensional parametric plot (Re $E_p(k)$, Im $E_p(k)$, $k$)~\citep{hu_knots_2021,wang_topological_2021}, which indicates an unlink structure for the trivial and nontrivial gapped insulating phases and an unknot for all the composite phases of the quadripartite model in Appendix~\ref{braid}.

\section{Phase boundaries}
\label{pb}

\begin{figure}[h!]
\centering
\includegraphics[scale=0.335]{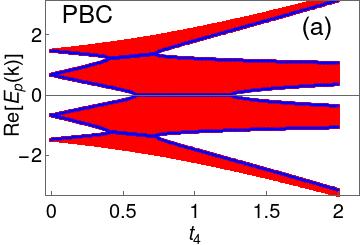}
\includegraphics[scale=0.335]{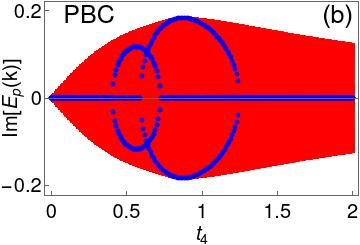}\\
\includegraphics[scale=0.335]{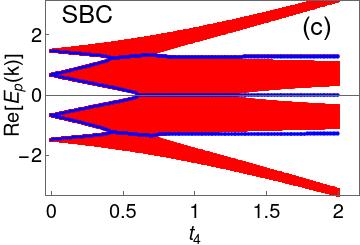}
\includegraphics[scale=0.335]{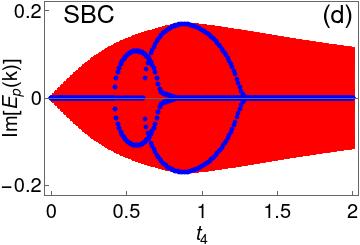}
\caption{The real and imaginary energy spectra $E_p(k)$ for $k\in[0,2\pi)$ of the quadripartite non-Hermitian SSH model with (a,b) periodic ($\eta=1$) and (c,d) special ($\eta=1/L$) boundary conditions. The blue lines show the energy responsible for the different band closing and opening with varying $t_4$.} 
\label{fig:phase}
\end{figure}

We now explain how to find the boundaries between different insulating and metallic phases. 
 The emergence of insulating phase $CI_m$ occurs by tuning the system parameters (particularly $t_4$) such that $\alpha^2-\beta|_{k=\pi}=0$. This leads to a quartic equation of $t_4$:
\begin{align}
e^{2\theta}t_4^4+2e^{\theta}(2t^2\hspace{-2pt}-\hspace{-2pt}t_2^2)t_4^2-4t^2t_2(1+e^{\theta})t_4+t_2^2(t_2^2+4t^2)=0,
\label{eq:16}
\end{align} 
which gives two solutions of our interest for $t_4$ as
\begin{align}
t_{m_{1,2}}\hspace{-3pt}:=\hspace{-1pt}e^{-\frac{\theta}{2}}\hspace{-3pt}\Bigg[\hspace{-2pt}\sqrt{\Gamma^2+t_2^2}\mp\hspace{-2pt}\sqrt{\frac{2t^2t_2\cosh\frac{\theta}{2}}{\sqrt{\Gamma^2+t_2^2}}\hspace{-2pt}-\hspace{-2pt}(\Gamma^2+2t^2)}\Bigg]\hspace{-1pt},\hspace{-2pt}
\label{eq:17}
\end{align}
where $\Gamma=\sqrt[3]{\lambda+\sqrt{\lambda^2+\xi^3}}+\sqrt[3]{\lambda-\sqrt{\lambda^2+\xi^3}}$, $\lambda=\frac{1}{2}t^2t_2\sinh (\theta/2)$ and $\xi=(t^2+t_2^2)/3$. 
These values of $t_{m_{1,2}}$ define the boundaries for two composite loops in the quadripartite model. The length of the minor axis of the first elliptical boundary in the imaginary spectrum of the model (blue line in Figs.~\ref{fig:phase}(b,d)) imparts a distance between these two points, i.e., $|t_{m_2}-t_{m_1}|$. 
Applying the band-touching condition $\beta|_{k=0}=0$, we obtain two more solutions of $t_4$ defined as  $t_{z_{1,2}}:=\frac{t^2e^{-\theta}}{t_2},\frac{t^2}{t_2}$, which are two boundaries of the other composite insulating phase $CI_z$.  The touching of bands and the phase boundaries are shown in Figs.~\ref{fig:phase}(a,b), where the energy spectra under periodic boundary conditions ($\eta=1$) display a clear bulk-boundary correspondence with the energy spectra in Figs.~\ref{fig:phase}(c,d) including zero modes and mid-gap states obtained using special boundary conditions ($\eta=1/L$)~\citep{vyas_topological_2021}.
 
The bi-orthonormality ($\langle \phi_{i}(k)|\psi_j(k)\rangle=\delta_{i,j}$) of the left and right eigenvectors breaks down when $N_p(k)=0$ for $p=1,2,3,4$, which results in non-diagonalisability of the non-Hermitian system at three special energies creating a coalescence of eigenvectors. Two of such special energies when considered for relevant bands at proper $k$, e.g., $E_{p=1,2 ~{\rm or}~ 3,4}(k=\pi)=\pm\sqrt{(t_{1r}t_{1l}+t_{2r}t_{2l}+t_{3r}t_{3l}+t_{4r}t_{4l})/{2}}$ and $E_{p=2,3}(k=0)=0$, lead to conditions identical to  $\alpha^2-\beta|_{k=\pi}=0$ and $\beta|_{k=0}=0$, respectively. Similarly, we compare the third special energy with the relevant bands (e.g., for $p=1,4$ and $p=2,3$) to get an equation as $E_{p}(k=\pi)=\pm\sqrt{t_{1r}t_{1l}+t_{2r}t_{2l}}$, whose solutions for $t_4$ give two hybrid points (HPs); they are $\tilde{t}_{m_{1,2}}=t_2e^{-\theta},t_2$ for our simplified choice of parameters (Appendix~\ref{cHP}). These HPs redefine the boundaries of the composite phases by introducing two new phases: (a) first phase for $t_4\in(\tilde{t}_{m_1},t_{m_1})$ is represented by separate loops of individual bands and these loops are characterized by quantized Zak phase ($\omega_{1,4}=\pi, \omega_{2,3}=0$) since the real energy of the bands are not degenerate at this HP boundary, and (b) second phase for $t_4\in(t_{m_2},\tilde{t}_{m_2})$ looks identical to $CI_z$ due to overlap with the boundary of $CI_z$. Moreover, two composite insulating phases $CI_m$ and $CI_z$ can merge together to form the composite metallic phase $CM_3$ when the system specifications are tuned properly, as shown in Figs.~\ref{fig:phase}(a,b). Apart from $\theta$ and $t_4$, $t_2$ controls mixing and ordering of $CI_m$ and $CI_z$. The boundaries of $CM_3$ depend on the overlap (separation) of the boundaries of $CI_m$ and $CI_z$ as shown in Fig.~\ref{fig:phase}.

\section{Geometric phase for composite loops}
\label{geo}
Finally, we discuss a systematic framework to describe the topology of the composite phases using extended periodicity of the composite loops in Fig.~\ref{fig:intro}. Since the bands participating in these composite loops are merged into one another in such a way that the system can be continuously changed to go from one band to another at the EPs~\citep{vyas_topological_2021}, the adiabatic cyclic evolution of a quantum state over the first Brillouin zone for calculating geometric phase can not be considered \footnote{Since $\displaystyle\lim_{\delta \to 0}[\phi_{p}(k_{EP}-\delta).\psi_{p}(k_{EP})]=0$ for $p^{th}$ band participating in the composite loop, and $k_{EP}$ is momentum at EPs.}. Nevertheless, the concept of geometric phase, which can purely be defined kinematically in terms of a cyclic overlap~\citep{pancharatnam_generalized_1956,berry_adiabatic_1987, 
anandan_geometric_1992, vyas2019pancharatnamzak} of contributing non-orthogonal vectors, still works for such composite loops \cite{wu_topology_2021}. Thus, we define a bi-orthonormal complex geometric (Zak) phase over a cyclic contour on the composite loop made of $n$ bands denoted by $p_1,p_2\dots p_n$: 
\begin{align}
&\omega_{p_1p_2\dots p_n}=i\displaystyle\oint \phi_{p_j}(k).\partial_k\psi_{p_j}(k)\; dk, \nonumber\\
&=i\displaystyle\int_0^{2\pi}\phi_{p_1}(k).\partial_k\psi_{p_1}(k)\; dk+i\displaystyle\int_{2\pi}^{4\pi}\phi_{p_2}(k).\partial_k\psi_{p_2}(k)\; dk \nonumber \\&+\dots+i\displaystyle\int_{2(n-1)\pi}^{2n\pi}\phi_{p_n}(k).\partial_k\psi_{p_n}(k)\; dk. \label{compZak}
\end{align}
Here we have fixed Arg$[\displaystyle\lim_{\delta \to 0}[\phi_{p}(k_{EP}-\delta).\psi_{p^\prime}(k_{EP})]]=0$ by suitable choice of the bi-orthonormal eigenvectors where $k_{EP}$ is momentum at EPs, and $p,p^\prime$ are consecutive bands on the composite loop. 
We can again write a gauge-invariant form as $\tilde{\omega}_{p_1p_2\dots p_n}=\omega_{p_1p_2\dots p_n}(\text{mod}\;2\pi)$.

\begin{figure}[h!]
\centering
\hspace{-15pt}\includegraphics[scale=0.675]{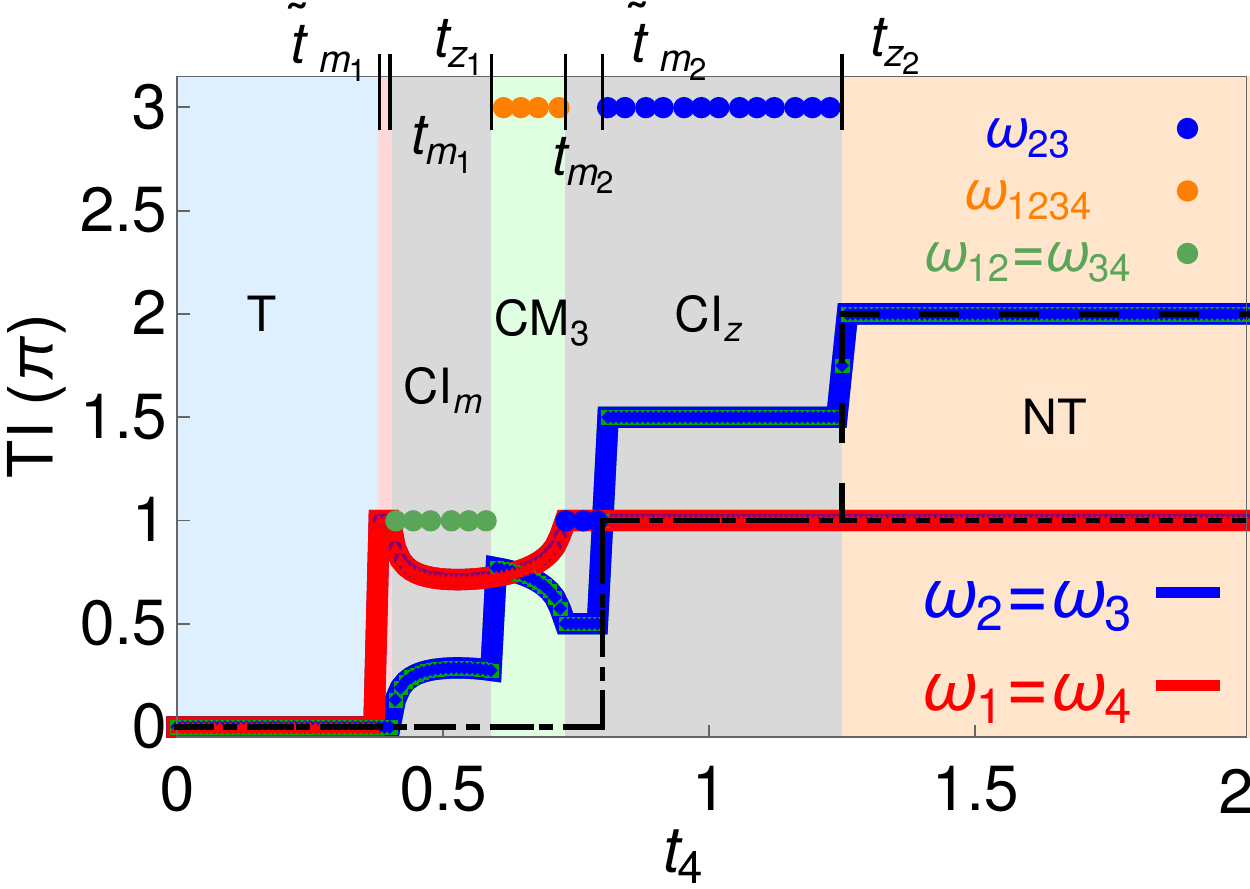}
\setlength{\belowcaptionskip}{-12pt}
\caption{ The topological invariants (TIs), $\omega_p, \omega_{p_1p_2\dots p_n}$,  for relevant energy bands of the above model. Different topological phases are  represented by different colors, such as trivial ($T$) phase by light blue color, composite insulating ($CI_m, CI_z$) phases by light grey color, composite metallic ($CM_3$) phase by light green color, nontrivial ($NT$) phase by light orange and hybrid region by light red color. The short and long dashed lines indicate the Zak phase in the Hermitian limit ($\theta=0$) for $p=1,4$ and $p=2,3$, respectively. The dotted lines denote $\omega_{p_1p_2\dots p_n}$ of the composite loops made of multiple bands in the composite insulating and metallic phases. The value of $\omega_p$ (Eq.~\ref{eq:5}) for individual bands in the composite phases indicates the contribution from the smooth contour by excluding the jump(s) at the exceptional point(s). 
The parameters are $t=1.0$, $t_2=0.8$, $\theta=0.75$.} 
\label{fig:phase1}
\end{figure}

 Applying the formula in Eq.~\ref{compZak} for composite loops in Fig.~\ref{fig:intro}(b), we find $\omega_{12}=\omega_{34}=\pi$ which are quantized indicating nontrivial topology of $CI_m$. We remind that the values of Zak phase of individual bands at this parameter regime are non-quantized due to the presence of EPs prohibiting generation of a unique cyclic path of evolution for individual band. Thus, $\omega_{p_1p_2\dots p_n}$ (or $\tilde{\omega}_{p_1p_2\dots p_n}$) acts a good topological invariant for the composite phases. We further emphasize that the contributing bands for the quantization of the geometric phase can only be determined by the cyclic loops formed on the parametric space in Fig.~\ref{fig:intro}. Following the above prescription, we calculate the geometric phase $\omega_{1234}=3\pi~(\tilde{\omega}_{1234}=\pi)$ and $\omega_{23}=3\pi~(\tilde{\omega}_{23}=\pi)$ for the composite loops in Figs.~\ref{fig:intro}(c,d) representing $CM_3$ and $CI_z$, respectively. The values of geometric phase for composite loops acting as  topological invariant for different topological phases are shown in Fig.~\ref{fig:phase1}. 

When we choose $t_1t_3<t_2^2$, the order of appearance of $CI_m$ and $CI_z$ reverses with increasing $t_4$. Surprisingly, the $CI_m$ phase is topologically trivial for such a parameter set although the composite loops still encircle a single EP discussed in Appendix~\ref{ctopo}. This is in contrast to the nontrivial topology of $CI_m$ phase when $t_1t_3>t_2^2$ discussed in the earlier parts. Therefore, while we notice that topology of the composite phase is generally nontrivial for associated composite loop encircling odd number of EPs (e.g., in bipartite and quadripartite SSH model in Fig.~\ref{fig:intro}) and is trivial for related composite loop encircling even number of EPs as in tripartite non-Hermitian SSH model, this is not a norm since we also have exceptions as for $CI_m$ phase with $t_1t_3>t_2^2$. 
\section{Conclusion}
\label{con}
In conclusion, we have shown that both adiabatic and nonadiabatic/dynamical descriptions are required in order to understand the topology of the phases in non-Hermitian multipartite 1D systems. While the topology of some phases with Hermitian counterparts can be understood invoking the concept of adiabatic cyclic evolution over individual bands, we surely need a more general concept of nonadiabatic cyclic evolution over the composite loop of multiple bands for describing the topology of phases without Hermitian counterparts. We further find that topology of the composite phases can be trivial or nontrivial. While we have here mostly explored topological classes AI or D$^\dagger$ with $\mathcal{S}_+$ sublattice symmetry of the non-Hermitian symmetry classes (Appendix~\ref{symmetry}), it is important to check whether the present formulation of cyclic nonadiabatic geometric phase based on composite loops on complex-energy parametric space applies to other symmetry classes of non-Hermitian systems! It seems to work for parity-time (PT) symmetric non-Hermitian models~\citep{kunst_biorthogonal_2018, wu_topology_2021,zhu_mathcalpt_2014,liu_metrology_2016,midya_non-hermitian_2020}. Finally, more studies are required to explore properties of the states represented by composite loops in 1D as well as their analogs in higher-dimensional systems~\citep{yuce_topological_2019,zou_observation_2021,du_effects_2021,zhang_topological_2021, Hengyun_Zhou2018}. The physical differences between various topologically trivial and nontrivial composite phases might be possible to explore by investigating dynamical properties of the models and dynamical encircling of EPs \cite{zhang_dynamically_2019}. We envision experimental probing of topology (geometric phase) of the states associated with composite loops in various controlled systems following the real Zak phase studies~\citep{atala_direct_2013,xiao_surface_2014,xiao_geometric_2015}.

\section{Acknowledgment}
We thanks Vivek Vyas and Neeraj Dhanwani for helpful discussions. This research is funded by the Ministry of Electronics $\&$ Information Technology (MeitY), India under the grant for ``Center for Excellence in Quantum Technologies" with Ref. No. 4(7)/2020-ITEA.   

\appendix
\section{\label{nssh2}nSSH2 Model}
\begin{figure}[h!]
\centering
\includegraphics[width=\linewidth]{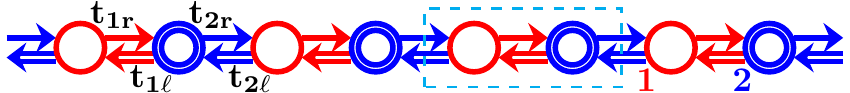}
\caption{A schematic of the bipartite non-Hermitian SSH model. The right moving hopping from $\sigma^{th}$ sub-unit site to the upcoming right site is $t_{\sigma r}$ (arrow) and left moving hopping to $\sigma^{th}$ sub-unit site from the upcoming right site is $t_{\sigma l}$ (double arrow), where $\sigma=1,2$ represents the sub-unit sites $1,2$.}
\label{fig:1}
\end{figure}

In this appendix, we describe topological properties of bipartite non-Hermitian SSH (nSSH2)  model~\citep{vyas_topological_2021,yin_geometrical_2018,shen_topological_2018,fu_extended_2020}  in Fig.~\ref{fig:1}.
The Hamiltonian of the model of $L$ unit cells (by using $s=2$, $\eta=0$ in Eq.~\ref{eq:1}) reads as
\begin{align}
H_{2}=\displaystyle\sum_{j=1}^{L-1}(t_{1r}c^\dagger_{j,2}c_{j,1}+t_{1l}c_{j,1}^\dagger c_{j,2})&+(t_{2r}c^\dagger_{j+1,1}c_{j,2}+\nonumber\\&t_{2l}c^\dagger_{j,2}c_{j+1,1}),
\label{eqa:1}
\end{align}
where $c^\dagger_{j,\sigma}(c_{j,\sigma})$ is fermionic creation (annihilation) operator on sublattice site $\sigma=1,2$ of the $j^{th}$ unit cell of the chain. Here,  $t_{\sigma r}$ is the right moving hopping from $\sigma^{th}$ sub-unit site to the right-next site and $t_{\sigma l}$ is the left moving hopping from right-next site to the $\sigma^{th}$ sub-unit site. Despite unequal directional hoppings $(t_{\sigma r} \ne t_{\sigma l})$, the presence of the translational symmetry for periodic boundary condition (PBC) ensures Fourier transformation which leads to the Hamiltonian in the momentum space:
\begin{align}
\hspace{-6pt}H_2=\hspace{-2pt}\displaystyle\sum_{k}[c^\dagger_{k,1},c^\dagger_{k,2}]\hspace{-2pt}\begin{bmatrix}
0&t_{1l}+t_{2r}e^{ik}\\
t_{1r}+t_{2l}e^{-ik}&0
\end{bmatrix}\hspace{-4pt}
\begin{bmatrix}
c_{k,1}\\c_{k,2}
\end{bmatrix}\hspace{-2pt},
\label{eqa:2}\hspace{-2pt}
\end{align}
with the momentum $k=2\pi j/L$ for $j=1,2,\hdots,L$. 
From Eq.~\ref{eqa:2}, we get two complex energy spectra for two energy bands of the chain as
 \begin{align}
 E_p(k)=\pm\sqrt{(t_{1r}+t_{2l}e^{-ik})(t_{1l}+t_{2r}e^{ik})},\;p=+,-.
\label{eqa:3} 
\end{align}
We have $E_+(k=\pi)=E_-(k=\pi)=0$ at two limits, i.e., $t_{1l}=t_{2r}$ or $t_{1r}=t_{2l}$. We set our parameters as $t_{1l}=t_{1r}=t$, $t_{2l}=t_2e^{\theta}$ and $t_{2r}=t_2$ in the following discussion where the non-Hermiticity is controlled by $\theta$. 

\begin{figure}[h!]
\centering
\includegraphics[scale=0.45]{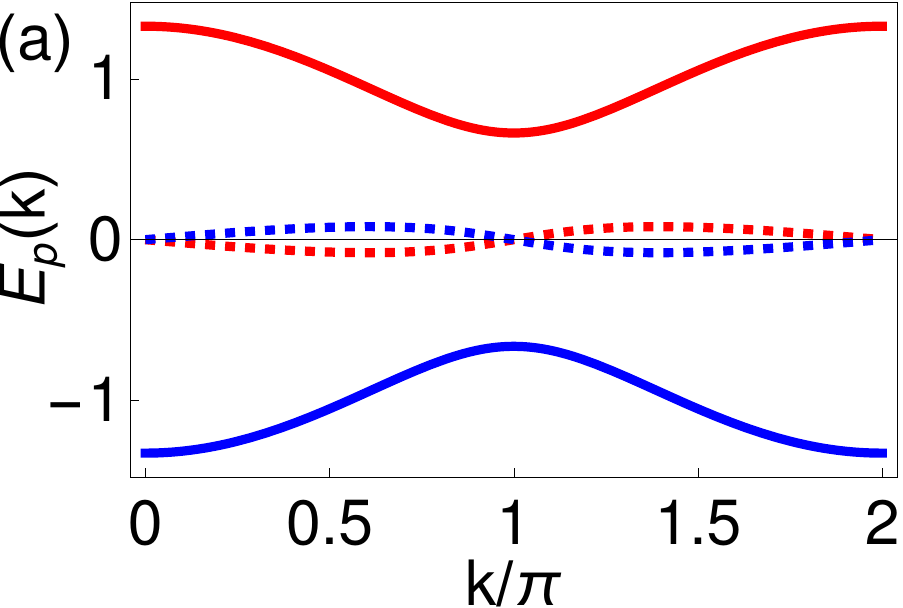}
\includegraphics[scale=0.45]{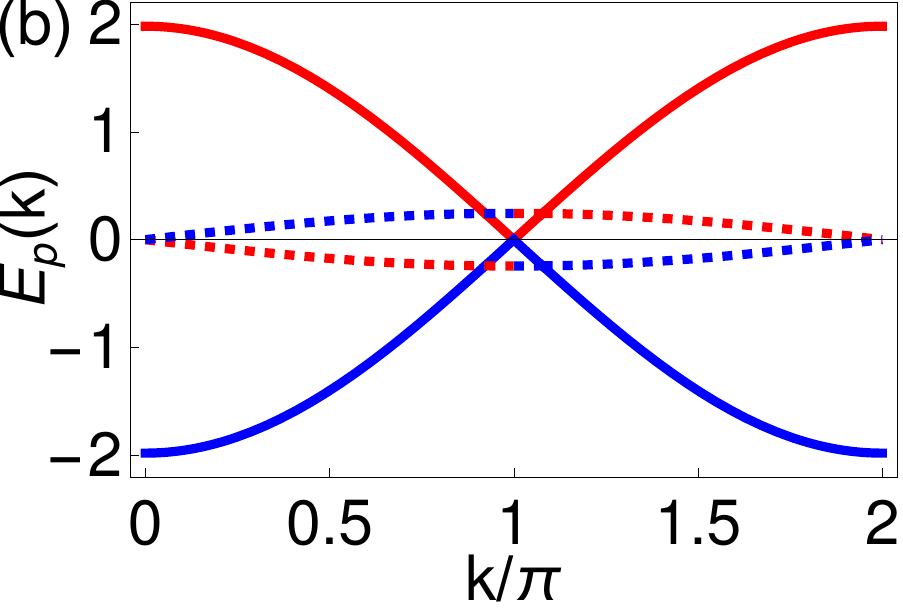}
\includegraphics[scale=0.45]{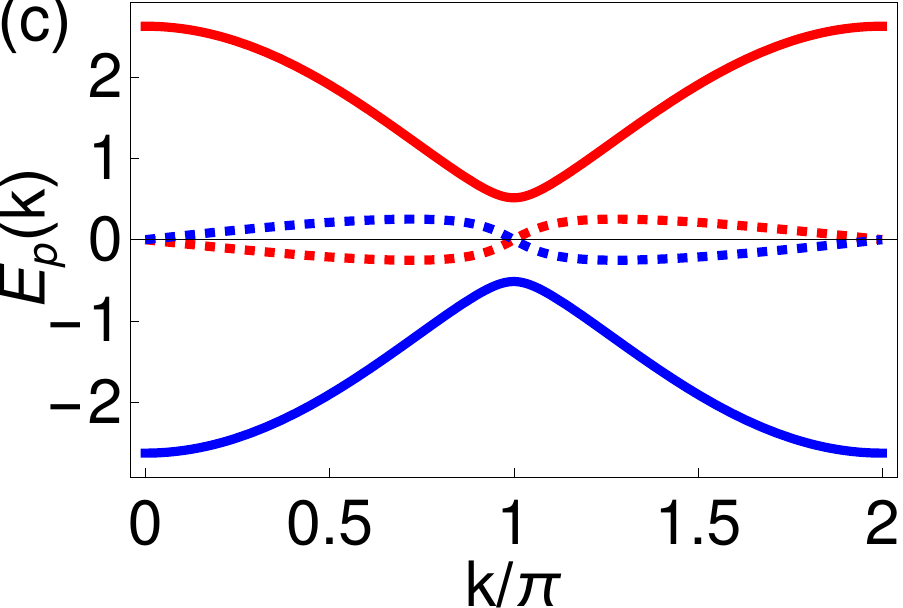}
\caption{The energy spectra of the nSSH2 model where the upper ($E_+(k)$) and the lower ($E_-(k)$) energy band are denoted using red and blue lines, respectively. The solid (dotted) line depicts the real (imaginary) part of the spectrum. The parameters are $t=1.0, \theta=0.5$ in all plots and (a) $t_2=0.25$ (trivial insulator), (b) $t_2=0.75$ (composite metal), and (c) $t_2=1.25$ (nontrivial insulator).}
\label{fig:2}
\end{figure}
The energy spectra of the two bands of the nSSH2 chain are shown in Fig.~\ref{fig:2} for different values of $t_{2}$.
From Eq.~\ref{eqa:3}, we obtain the complex energy bandgap, $\Delta E=2\sqrt{(t-t_2)(t-t_2e^{\theta})}$. Similar to the Hermitian case, the nSSH2 chain possesses a real energy gap in Figs.~\ref{fig:2}(a,c) for $t_2<te^{-\theta}$ and $t_2>t$. These gapped phases are regarded as topologically trivial ($T$) and nontrivial ($NT$) insulators following the nomenclature in the Hermitian limit. Further, the real energy bandgap vanishes between $t_2=te^{-\theta}$ and $t_2=t$ at $k=\pi$ in Fig.~\ref{fig:2}(b), which generates a new metallic phase without any Hermitian counterpart \cite{vyas_topological_2021}.

\begin{figure}[h!]
\includegraphics[width=\linewidth]{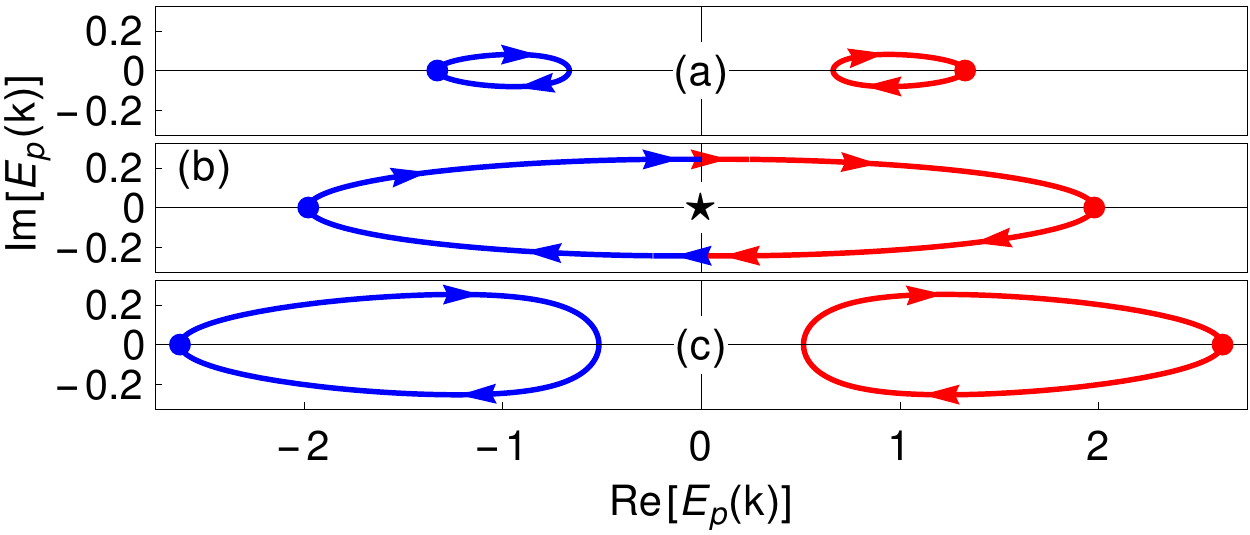}
\caption{The complex energy spectra of nSSH2 model on the parametric space of Re[$E_p(k)$] and Im[$E_p(k)$] for $t=1.0, \theta=0.5$ in all plots and (a) $t_2=0.25$ (trivial insulator), (b) $t_2=0.75$ (composite metal), and (c) $t_2=1.25$ (nontrivial insulator). The red and blue color denote the upper ($E_+(k)$) and the lower ($E_-(k)$) band, respectively. The dots show the energy bands at $k=0$ and the arrows indicate the direction of complex energy loops as $k$ changes from $0$ to $2\pi$.}\label{fig:3}
\end{figure}

The left ($\phi_p$) and right ($\psi_p$) eigenvectors of the non-Hermitian system are obtained from the $k$-space Hamiltonian in Eq.~\ref{eqa:2}.
\begin{align}
\phi_p(k)=\frac{1}{\sqrt{2}}\begin{bmatrix}
1&\frac{E_p(k)}{t_{1r}+t_{2l}e^{-ik}} 
\end{bmatrix},
\psi_{p}(k)=\frac{1}{\sqrt{2}}\begin{bmatrix}
1\\
\frac{E_p(k)}{t_{1l}+t_{2r}e^{ik}}
\end{bmatrix}.
\label{eqa:4}
\end{align}
These eigenvectors follow the bi-orthonormality relation $\langle\phi_{p}(k)|\psi_{p'}(k)\rangle=\delta_{p,p'}$. As discussed earlier, both the real and imaginary parts of the two energy bands at $k=\pm\pi$ vanish when $t_{2l}=t_{1r}$ or $t_{2r}=t_{1l}$. These special points are called exceptional points (EPs) (shown by a ``star" in Fig.~\ref{fig:3}(b)) at which the eigenenergies coalescence and the eigenfunctions become self-orthogonal~\citep{ryu_analysis_2012,ding_emergence_2016,heiss_chirality_2001,wu_topology_2021}, which leads to a non-diagonalizable Hamiltonian. 

We now  illustrate the complex energy spectra from Eq.~\ref{eqa:3} on the parametric  space of Re[$E_p(k)$] and Im[$E_p(k)$] in Fig.~\ref{fig:3}. 
 In the trivial and nontrivial insulating phase, each of the complex energy band traces a cyclic elliptical boundary, and they are separated by some real energy  in Figs.~\ref{fig:3}(a,c). Such loops are formed by continuous real and imaginary spectra of the system, along with a real energy gap indicated by the distance between the nearest vertices of the two ellipses. A composite loop in Fig.~\ref{fig:3}(b) is formed by two bands together enclosing one EP at $k=\pi$. This composite loop represents a composite metallic phase $CM_1$ where the subscript denotes one EP.   
\begin{figure}[h!]
\centering
\hspace{-15pt}\includegraphics[width=\linewidth]{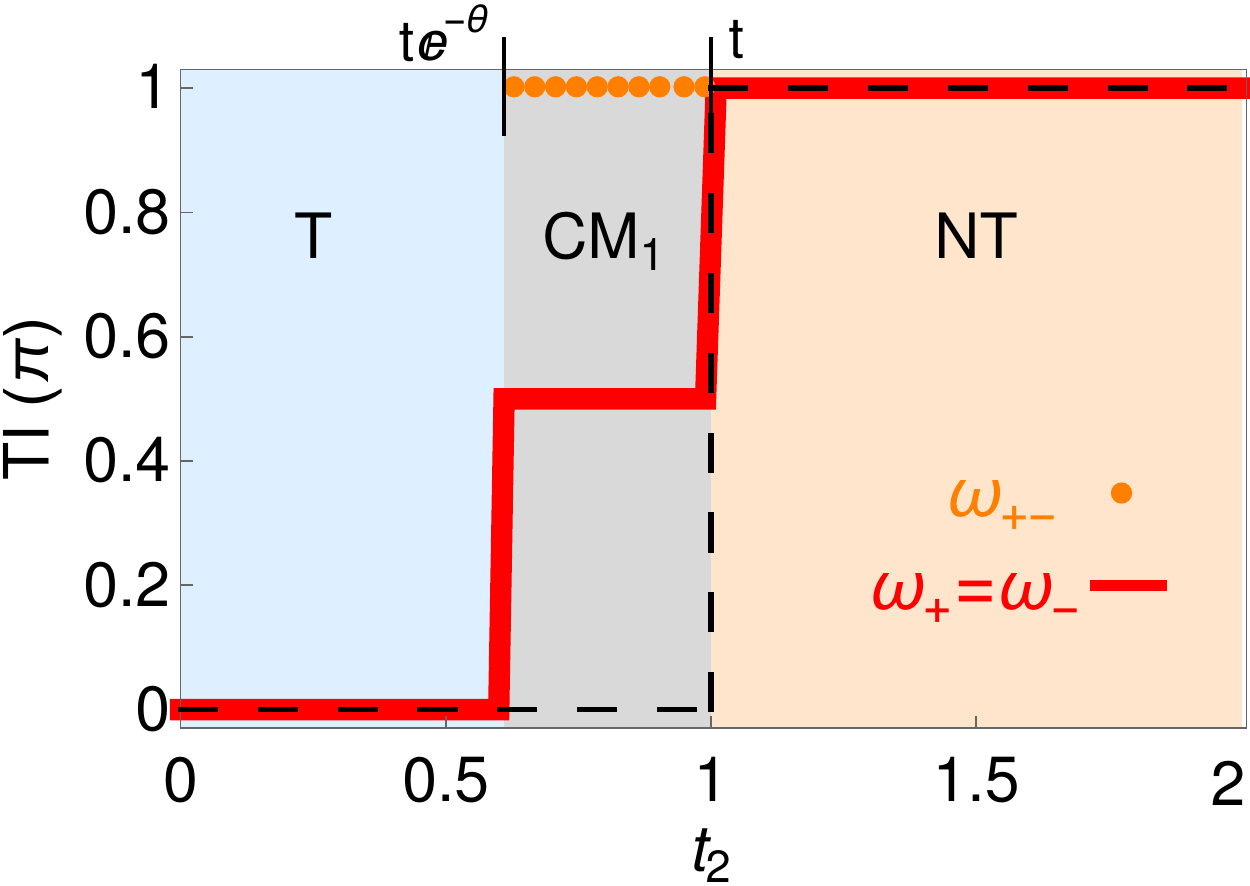}\setlength{\belowcaptionskip}{-5pt}
\caption{Topological invariant (TI), $\omega_{\pm}, \omega_{+-}$, for complex energy bands of nSSH2 model with inter-cell hopping $t_2$. The parameters are $t=1.0, \theta=0.5$. Different colors denote the trivial ($T$) phase (light blue) for $t_2\in(0,e^{-\theta})$, composite metallic ($CM_1$) phase (light gray) for $t_2\in(e^{-\theta},1)$ and nontrivial ($NT$) phase (light orange) for $t_2>1$. The dashed line gives the Zak phase for both bands in the Hermitian limit ($\theta=0$). The dotted line in the $CM_1$ phase denotes the composite Zak phase ($\omega_{+-}$).}\label{fig:5}
\end{figure}

The concepts of bi-orthonormal complex geometric (Zak) phase as topological invariant are applied in the main text for a cyclic adiabatic evolution over each separate loop (Eq.~\ref{eq:5}) and for nonadiabatic evolution over a cyclic contour on the composite loop (Eq.~\ref{compZak}) to study topology of the multipartite non-Hermitian systems. The complex Zak phase for separate loops in Figs.~\ref{fig:3}(a,c) is quantized, and its values for both bands are  $\omega_{\pm}=0,\pi$, respectively, for topologically trivial and nontrivial phase. The geometric phase $\omega_{+-}=\pi$ for the composite loop in  Fig.~\ref{fig:3}(b) is also quantized and non-zero, which indicates nontrivial topology of $CM_1$ phase. The complex bands together in the $CM_1$ phase form a M{\"o}bius strip topology as discussed earlier in Ref.~\cite{vyas_topological_2021}. We summarize the above discussion in Fig.~\ref{fig:5}, where $\omega_{\pm}$ for individual bands in the $CM_1$ phase denote the contribution from the smooth contour by excluding the amplitude of overlap of left and right eigenvectors at the EP.

 \section{\label{nssh3}nSSH3 model}
 
\begin{figure}[h!]
\centering
\includegraphics[width=\linewidth]{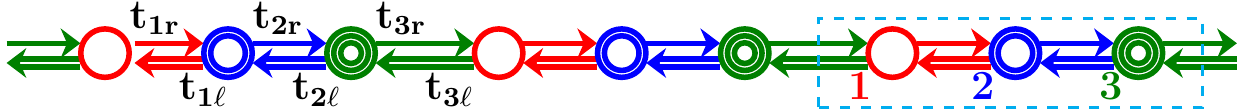}
\caption{The tripartite non-Hermitian SSH model with three sites per unit cell. The right moving hopping from $\sigma^{th}$ sub-unit site to the right-next site is $t_{\sigma r}$ (arrow) and the left moving hopping to the $\sigma^{th}$ sub-unit site from the right-next site is $t_{\sigma l}$ (double arrow), where $\sigma$ runs over the sub-units $1,2,3$.}
\label{fig:6}
\end{figure}

\begin{figure}[h!]
\centering
\hspace{-20pt}\includegraphics[width=0.5\linewidth]{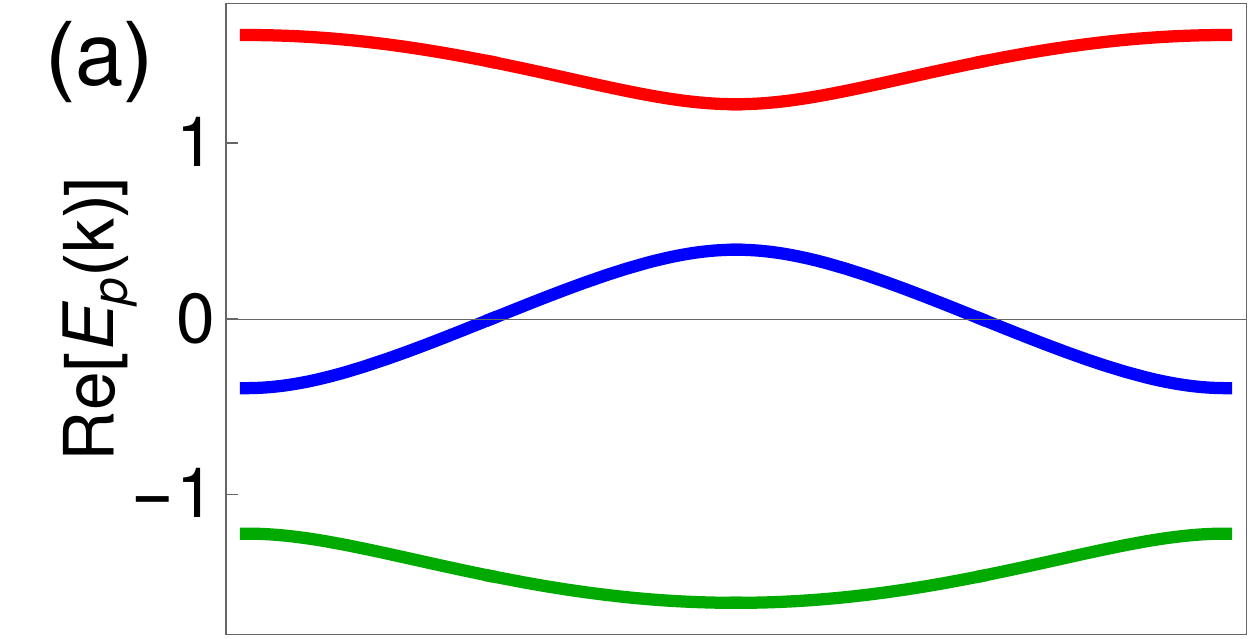}
\includegraphics[width=0.5\linewidth]{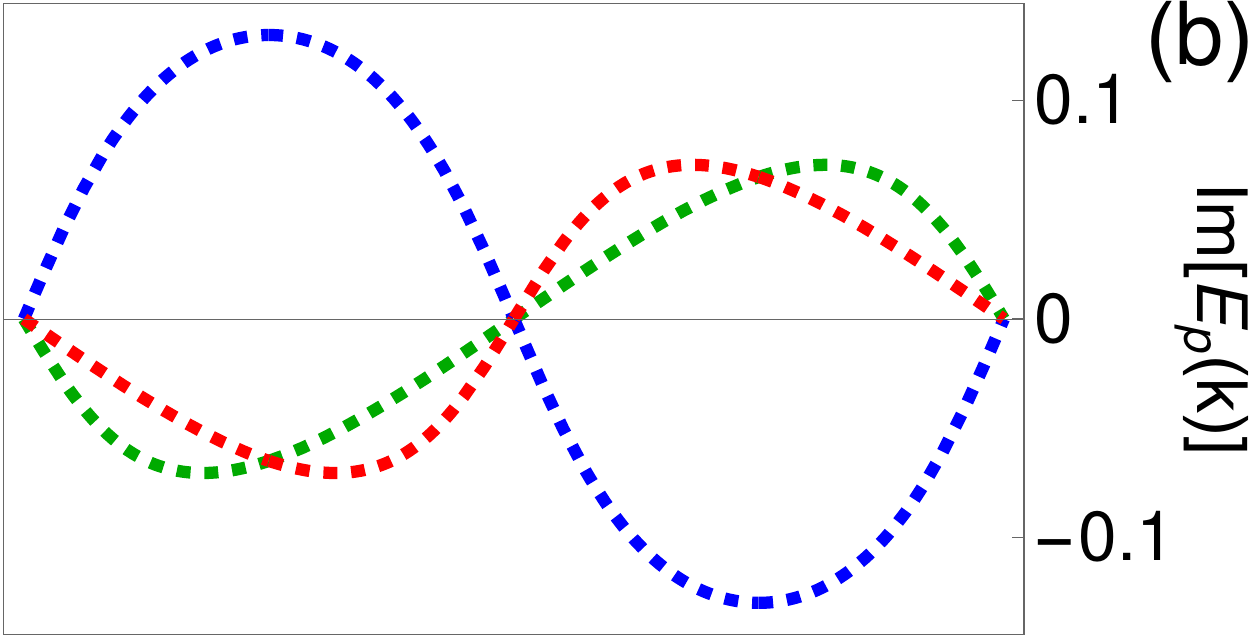}\\
\hspace{-20pt}\includegraphics[width=0.5\linewidth]{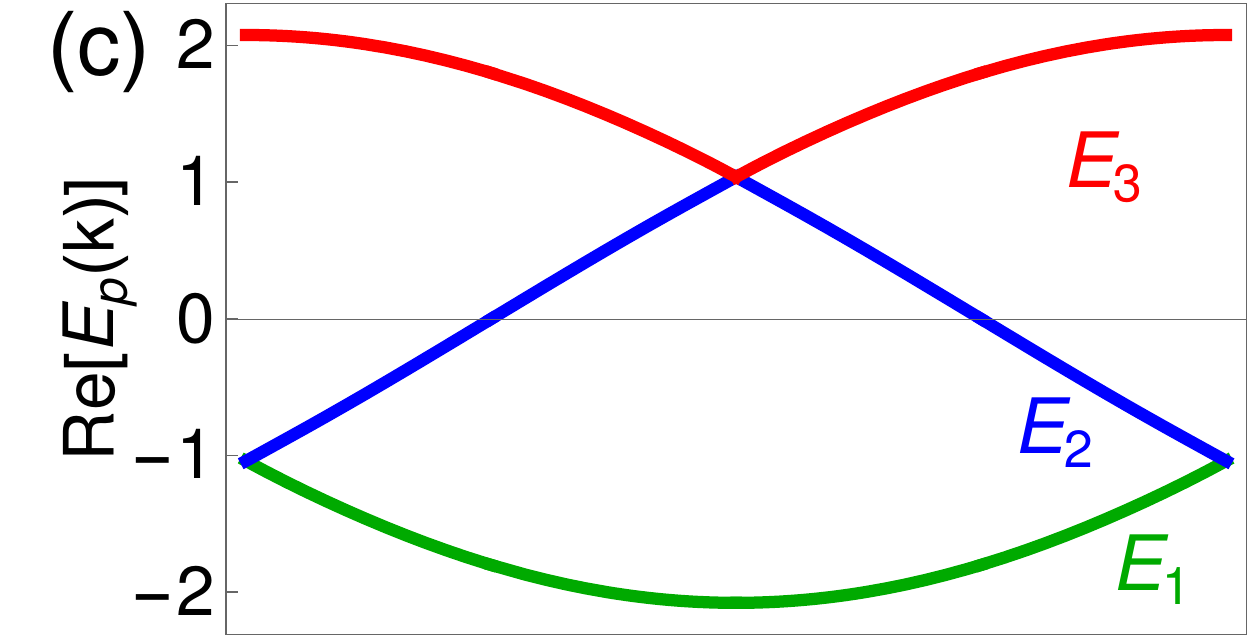}
\includegraphics[width=0.5\linewidth]{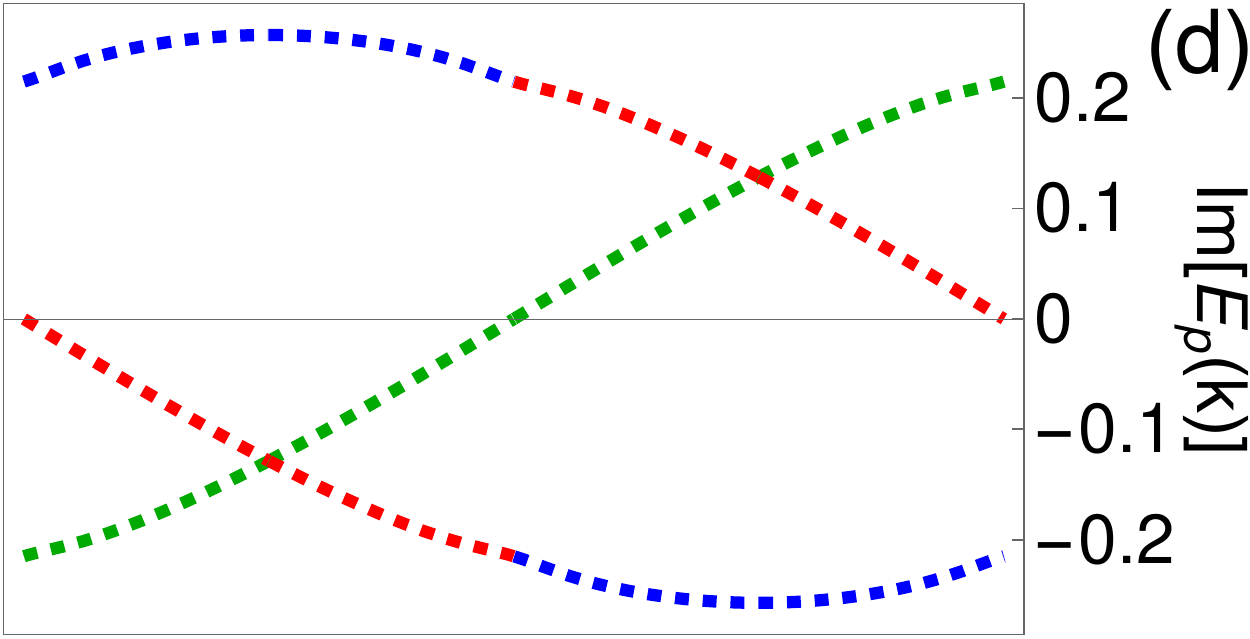}\\
\hspace{-20pt}\includegraphics[width=0.5\linewidth]{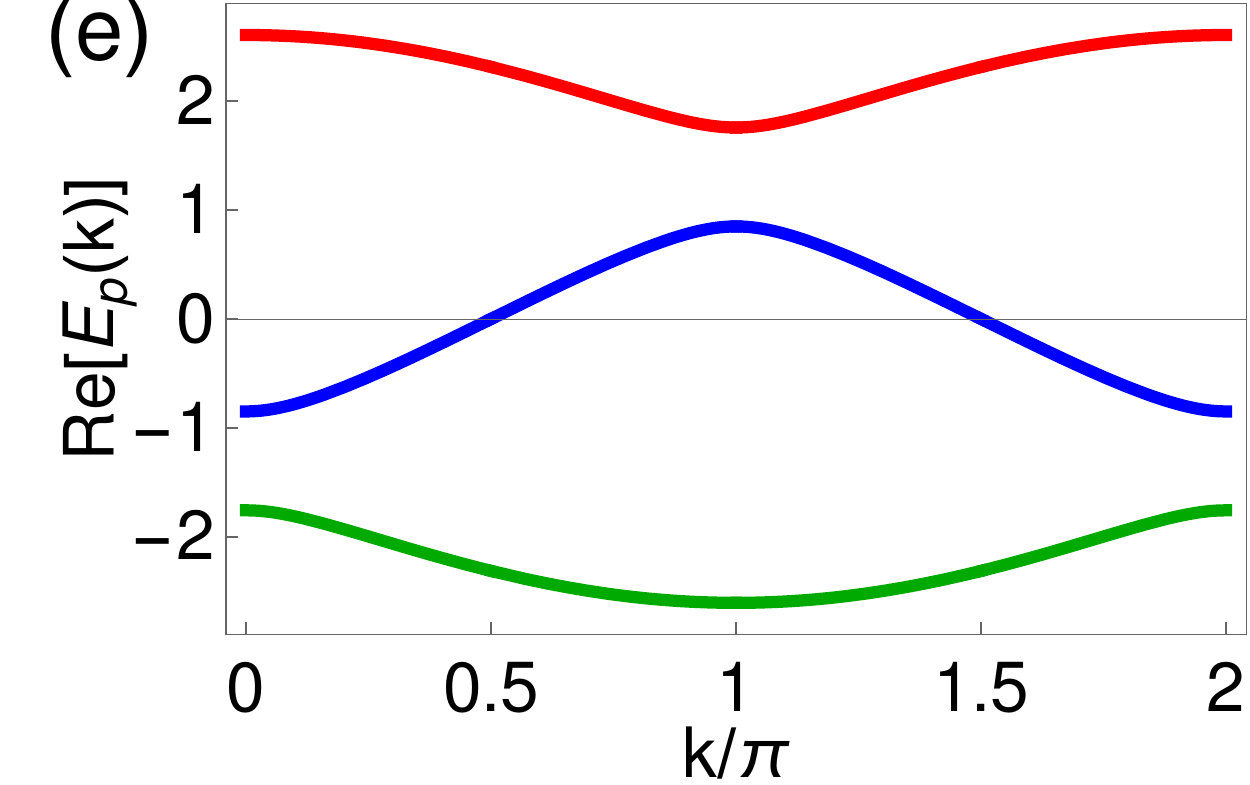}
\includegraphics[width=0.5\linewidth]{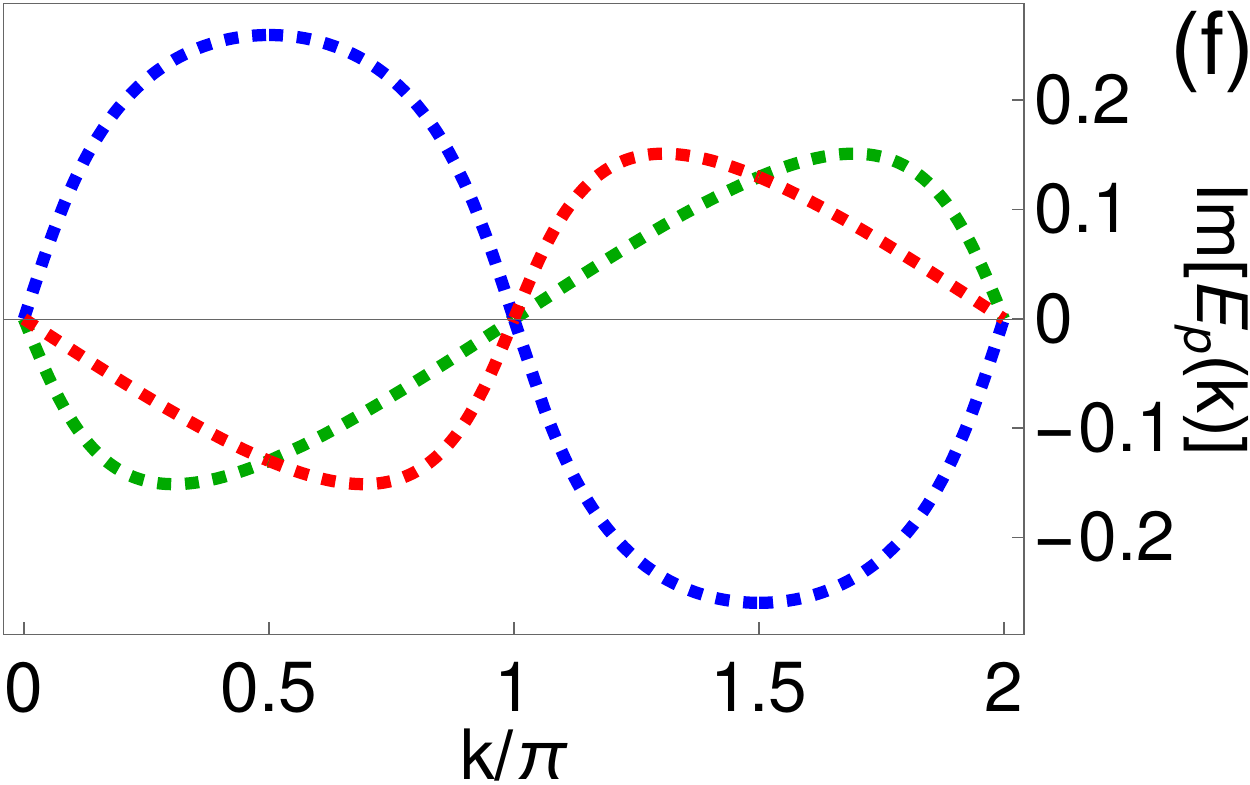}
\caption{The dispersion spectra of the nSSH3 model where three energy bands $E_1,\;E_2,\;E_3$ are represented by green, blue, red lines, respectively. The solid (dotted) lines depict the real (imaginary) part of the  spectrum. The parameters are  $t=1.0,\theta=0.75$ in all plots and (a,b) $t_3=0.25$ (trivial insulator), (c,d) $t_3=0.75$ (composite metal) and (e,f) $t_3=1.25$ (nontrivial insulator).}
\label{fig:7}
\end{figure}

 The topology of tripartite non-Hermitian SSH model (nSSH3) (Fig.~\ref{fig:6}) is highlighted in this appendix. 
 The Hamiltonian for nSSH3 model is
\begin{align}
H_{3}=\displaystyle\sum_{j=1}^{L-1}&(t_{1r}c^\dagger_{j,2}c_{j,1}+t_{1l}c_{j,1}^\dagger c_{j,2}+t_{2r}c^\dagger_{j,3}c_{j,2}+t_{2l}c^\dagger_{j,2}c_{j,3}\nonumber\\&+t_{3r}c^\dagger_{j+1,1}c_{j,3}+t_{3l}c^\dagger_{j,3}c_{j+1,3}),
\label{eqa:6}
\end{align}
where $c^\dagger_{j,\sigma}(c_{j,\sigma})$ is fermionic creation (annihilation) operator at sub-lattice site $\sigma=1,2,3$ of the $j^{th}$ unit cell of the chain. The right (left) moving hopping from (to) $\sigma^{th}$ sub-unit site to (from) the nearest right site is given by $t_{\sigma r}$ ($t_{\sigma l}$). 
The number of unit cell $L$ is restricted to an even value in order to avoid the extra zero energy state due to the presence of sublattice symmetry~\citep{sirker_boundary_2014,he_topology_2018,he_non-hermitian_2020} in the chain.
The translational invariant Hamiltonian (Eq.~\ref{eqa:6}) with PBC after a Fourier transformation in the momentum space reads as
\begin{align}
\hspace{-7pt}H_3=\displaystyle\sum_{k}[c^\dagger_{k,1},c^\dagger_{k,2},c^\dagger_{k,3}]\hspace{-2pt}\begin{bmatrix}
0&t_{1l}&t_{3r}e^{ik}\\
t_{1r}&0&t_{2l}\\
t_{3l}e^{-ik}&t_{2r}&0
\end{bmatrix}\hspace{-3pt}
\begin{bmatrix}
c_{k,1}\\c_{k,2}\\c_{k,3}
\end{bmatrix}\hspace{-3pt},\hspace{-7pt}
\label{eqa:7}
\end{align}
where $k=\frac{2\pi j}{L}$ with $j=1,2,\hdots,L$ are momenta of the system. 
As a consequence of three sites per unit cell, the system possesses three complex energy bands~\citep{ryu_analysis_2012} given by relation
 \begin{align}
E_{p}(k)=e^{i\frac{2\pi p}{3}}{\sqrt[3]{w+\sqrt{w^2-v^3}}}+e^{-i\frac{2\pi p}{3}}{\sqrt[3]{w-\sqrt{w^2-v^3}}},
\label{eqa:8}
\end{align}
 where $v=(t_{1l}t_{1r}+t_{2l}t_{2r}+t_{3l}t_{3r})/3, w=(t_{1r}t_{2r}t_{3r}e^{ik}+t_{1l}t_{2l}t_{3l}e^{-ik})/2$ and the band index $p=1,2,3$. The degeneracies in the  complex energy bands, i.e., $E_{1}(0)=E_{2}(0)$ or $E_{2}(\pi)=E_{3}(\pi)$, provide the band closing condition when $[w^2-v^3]_{k=0,\pi}=0$. 
 We here particularly consider a set of parameters with intrinsic inversion symmetry in the unit cell, i.e., the center of the unit cell coincides with the axis of inversion in the unit cell, in other words, $t_{1l}=t_{2r}=t$, $t_{1r}=t_{2l}=t$. The non-Hermiticity is introduced through the inter-unit cell hopping as $t_{3l}=t_3e^{\theta}$ and $t_{3r}=t_3$, which break the overall inversion symmetry in the chain. 
The dispersion spectra of the model for the above set of the parameters are shown in Fig.~\ref{fig:7}. 

The real parts of complex energy spectra with energy gaps are shown in Figs.~\ref{fig:7}(a,e) for the topologically trivial ($T$) and nontrivial ($NT$) insulators, which are reminiscence of the corresponding Hermitian model. However, the corresponding imaginary parts of the energy spectra are  gapless and continuous in Figs.~\ref{fig:7}(b,f) due to the presence of non-Hermiticity. 
The degeneracies in the complex energies give rise to the EPs at energies $E_{1}(0)=E_{2}(0)$ and $E_{2}(\pi)=E_{3}(\pi)$. 
In Figs.~\ref{fig:7}(c,d), the real part of first ($E_1$) and second ($E_2$) energy band and that of second ($E_2$) and third ($E_3$) energy band touch each other at $k=0,2\pi$ and $k=\pi$, respectively. The imaginary parts of the energy bands show discontinuity at the corresponding $k$-values. These degeneracies in the real energies and discontinuities in the imaginary energies signal the crossing of EP.

The band closing condition, $[w^2-v^3]_{k=0,\pi}=0$,i.e.,
\begin{align}
27(t_{1r}t_{2r}t_{3r}+t_{1l}t_{2l}t_{3l})^2-4(t_{1l}t_{1r}+t_{2l}t_{2r}+t_{3l}t_{3r})^3=0,
\end{align}
 for the special choice of parameters leads to the condition: 
\begin{align}
\hspace{-5pt}e^{3\theta}t_3^6+24e^{2\theta}t^2t_3^4-3(9e^{2\theta}+2e^{\theta}+9)t^4t_3^2+32t^6\hspace{-2pt}=\hspace{-2pt}0,\hspace{-2pt}
\label{eqa:9}
\end{align}
which has only two possible real solutions for $t_3$, and they are
\begin{equation}
t_{m_{1,2}}=te^{-\frac{\theta}{2}}\sqrt{6\cosh\frac{\theta}{2}\cos\Big(\frac{\pi\pm\varphi}{3}\Big)-2};\;\tan\varphi=\sinh\frac{\theta}{2}.
\label{eqa:10}
\end{equation}
 The boundaries of the composite metallic phase encircling
two EPs ($CM_2$) are determined by $t_3\in(t_{m_1},t_{m_2})$.

 The left and right bi-orthonormal eigenvectors of the nSSH3 model for general parameters, respectively, are:
\begin{align}
\phi_p(k)=\begin{bmatrix}
\frac{E_p^2(k)-t_{2l}t_{2r}}{\sqrt{N_p}} &
\frac{E_p(k)t_{1l}+t_{2r}t_{3r}e^{ik}}{\sqrt{N_p}}&
\frac{E_p(k)t_{3r}e^{ik}+t_{1l}t_{2l}}{\sqrt{N_p}}
\end{bmatrix},
\label{eqa:11}
\end{align}
\vspace{-20pt}
\begin{align}
\psi_p(k)=\frac{1}{\sqrt{N_p}}\begin{bmatrix}
E_p^2(k)-t_{2l}t_{2r}\\
E_p(k)t_{1r}+t_{2l}t_{3l}e^{-ik}\\
E_p(k)t_{3l}e^{-ik}+t_{1r}t_{2r}
\end{bmatrix},
\label{eqa:12}
\end{align}
where $N_p(k)=(E_p^2(k)-t_{2l}t_{2r})(3E_p^2(k)-(t_{1l}t_{1r}+t_{2l}t_{2r}+t_{3l}t_{3r}))$ is the normalization constant for $p^{th}$ band. The bi-orthonormality relation is $\langle \phi_{p}(k).\psi_{p'}(k) \rangle=\delta_{p,p'}$ for band indices $p,p'$. This bi-orthonormality relation is meaningless when $N_p=0$ as it leads to the coalescence of eigenvectors and a defective Hamiltonian~\citep{ding_emergence_2016,wu_topology_2021,bergholtz_exceptional_2021}. The condition $N_p=0$ is satisfied when $E_p(k)=\pm\sqrt{(t_{1l}t_{1r}+t_{2l}t_{2r}+t_{3l}t_{3r})/{3}}$ or $E_p(k)=\pm\sqrt{t_{2l}t_{2r}}$. The first relation leads to  degeneracies for pair of bands at some particular $k$: $E_{1,2}(k=0)=-\sqrt{(t_{1l}t_{1r}+t_{2l}t_{2r}+t_{3l}t_{3r})/{3}}$ and $E_{2,3}(k=\pi)=\sqrt{(t_{1l}t_{1r}+t_{2l}t_{2r}+t_{3l}t_{3r})/{3}}$. The solutions of these degeneracy conditions in terms of $t_3$ are again $t_{m_1},t_{m_2}$, which are the EPs of the nSSH3 model as the vorticity~\citep{ghatak_new_2019,yin_geometrical_2018} of the related bands is half integer at these points $t_{m_1}$ and $t_{m_2}$ in Fig.~\ref{fig:vorticity} (Appendix~\ref{vorticity}). The second relation leads to $E_{1}(k=0)=-\sqrt{t_{2l}t_{2r}}$, $E_{3}(k=\pi)=\sqrt{t_{2l}t_{2r}}$ which give $t_3=te^{-\theta}$ and $E_{2}(k=0,\pi)=\mp\sqrt{t_{2l}t_{2r}}$ which give $t_3=t$ for the special choice of parameters. These values of $t_3$ are the special points (named as $\tilde{t}_{m_{1,2}}=t,te^{-\theta}$) of the nSSH3 chain~\citep{shen_topological_2018}. These points are called hybrid points (HPs) (Appendix~\ref{vorticity}) due to a zero vorticity and a defective Hamiltonian of the model at these points. 
We note that, there is no degeneracy associated with these two HPs, unlike the EPs in the chains. Finally, in the Hermitian limit, all four special points (2EPs, 2HPs) merge to a single trivial-to-topological phase transition point, i.e., $t_3=t$~\citep{shen_topological_2018}.

\begin{figure}[h!]
\centering
\hspace{-5pt}\includegraphics[width=\linewidth]{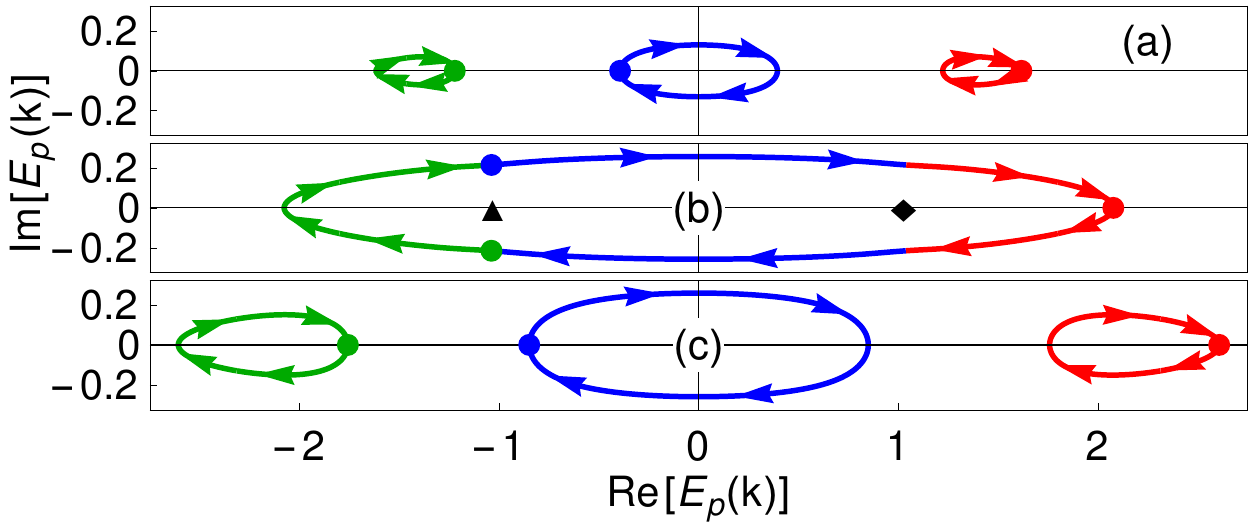}
\setlength{\belowcaptionskip}{-0pt}
\caption{The complex energy spectra of nSSH3 model on the parametric space of Re[$E_p(k)$] and Im[$E_p(k)$] for the parameters $t=1.0, \theta=0.75$ in all plots and (a) $t_3=0.25$ (trivial insulator), (b) $t_3=0.75$ (composite metal) and (c) $t_3=1.25$ (nontrivial insulator). The three energy bands $E_1(k),\;E_2(k),\;E_3(k)$ are denoted by green, blue, red lines, respectively. The dots show the energy band value at $k=0$, and the arrows indicate the direction of complex energy loops as $k$ changes from $0$ to $2\pi$.}\label{fig:8}
\end{figure}

The plots of complex energy spectra on the parametric plane of Re[$E_p(k)$] and Im[$E_p(k)$] are shown in Fig.~\ref{fig:8}.
Following the dispersion spectra in Figs.~\ref{fig:7}(a,b,e,f), we find that the individual energy bands form cyclic loops separated by some real energy gap in Figs.~\ref{fig:8}(a,c) in the trivial and nontrivial insulating phase of the nSSH3.
 Further, the Fig.~\ref{fig:8}(b) depicts a composite loop created by all three bands, which encloses two EPs ($\blacktriangle,\blacklozenge$). The composite loop represents a composite metallic phase named as $CM_2$.   
 The imaginary energies~\citep{ryu_analysis_2012,heiss_chirality_2008} and the eigenvectors~\citep{dembowski_encircling_2004,heiss_phases_1999} of two bands are exchanged while crossing these EPs.

\begin{figure}[h!]
\centering
\hspace{-15pt}\includegraphics[width=\linewidth]{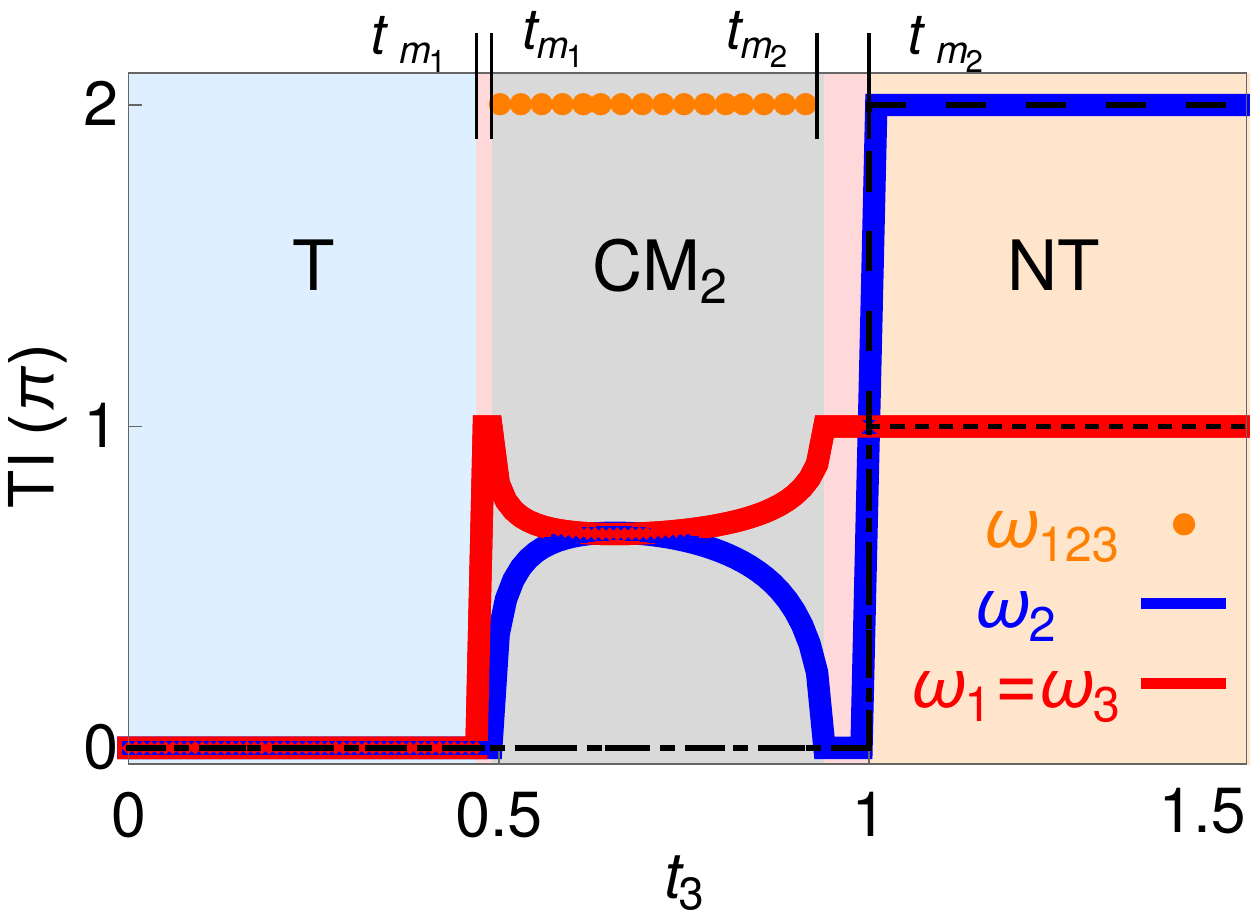}
\caption{Topological invariant (TI), $\omega_p,\;\omega_{p_1p_2\hdots p_n}$, for complex energy bands of nSSH3 model with inter-cell hopping $t_3$. The parameters are $t=1.0, \theta=0.75$. The model is in trivial ($T$) phase (light blue) for $t_3\in(0,\tilde{t}_{m_1})$, composite metallic ($CM_2$) phase (light grey) for $t_3\in(t_{m_1},t_{m_2})$, and nontrivial ($NT$) phase (light orange) for $t_3> \tilde{t}_{m_2}$. The short and long dashed lines indicate the Zak phase in
the Hermitian limit $(\theta=0)$ for $p=1,3$ and $p=2$, respectively. The dotted line shows the composite geometric phase ($\omega_{123}$) in the $CM_2$ phase. The light red regions indicate the insulating phases between the hybrid and exceptional points.}\label{fig:10}
\end{figure}
The topological phases of the Hermitian tripartite model are characterized by the Zak phase \cite{zak_berrys_1989}, which is quantized at either $0$ or $\pi$ for this model due to the inversion symmetry \cite{xiao_surface_2014}. We show the Zak phase values by dashed lines in Fig.~\ref{fig:10} where the value of $2\pi$ is equivalent to $0$ when the module $2\pi$ is applied. The topological invariant for the nSSH3 chain with increasing inter-cell hopping $t_3$ is shown in Fig.~\ref{fig:10}. The complex Zak phase for separate loops of individual bands in Figs.~\ref{fig:8}(a,c) is quantized and real for our sublattice symmetric tripartite model, and its values for all three bands is zero for $t_3<te^{-\theta}$ indicating a topologically trivial insulating phase. The complex Zak phase values for $t_3>t$ are $\omega_{1,3}=\pi$ and $\omega_{2}=2\pi~(\tilde{\omega}_2=0)$, which predict a topologically nontrivial insulating phase for the band $p=1,3$. In the $CM_2$ phase, 
the cyclic adiabaticity breaks down due to discontinuity at the EP. Therefore, the composite geometric phase (Eq.~\ref{compZak}) is used as a topological invariant which has a quantized value of $2\pi$ for $t_3\in(t_{m_1},t_{m_2})$. The composite Zak phase ($\omega_{123}$) over the composite loop (Fig.~\ref{fig:8}(b)) includes the contributions from the continuous part of the complex Zak phase of all the energy bands. The complex Zak phase of the individual bands, excluding the discontinuity at the EPs, shows a non-quantized value in the $CM_2$ phase in contrast to that in the $CM_1$ phase of the nSSH2 model. The complex energies at the band-closings are responsible for this non-quantization in the nSSH3 model, whereas the purely imaginary energies at the band-closings are responsible for the quantized complex Zak phase in the $CM_1$ phase of the nSSH2 model. Moreover, we emphasize that the $CM_2$ phase is non-topological since $\tilde{\omega}_{123}=0$.
Furthermore, the complex Zak phase between the HPs and EPs boundaries, i.e., $\tilde{t}_{m_1}< t_3< t_{m_1}$ and $t_{m_2}< t_3<\tilde{t}_{m_2}$ is also quantized with a value $\pi$ for the bands $p=1,3$ and zero for the band $p=2$ which represents  hybrid insulating ($HI$) phase shown by light red region in Fig.~\ref{fig:10}. The individual bands form separate loops in the $HI$ phase due to lack of degeneracy at these HPs.

\section{\label{nssh4}nSSH4 model}
In this appendix, we provide the details of our results on the Hermitian
and non-Hermitian quadripartite SSH model (nSSH4). We also include the complex energy spectra, other possible topological phases, and higher-order EPs for a different set of parameters in the quadripartite model. 
\subsection{\label{cHermitian}{}Topological phases in Hermitian limit}
The closing and reopening of energy bands with some parametric change in the Hamiltonian are used here to determine the topological phase transitions. In the Hermitian limit ($t_{\sigma r}=t_{\sigma l}=t_{\sigma}$), we find two band-closing conditions for the quadripartite SSH model from Eq.~\ref{eq:15}; these are (a) $E_1(\pi)=E_2(\pi)$ or $E_3(\pi)=E_4(\pi)$ and (b) $E_2(0)=E_3(0)$. The first condition leads to the equation:  
\begin{align}
\hspace{-10pt}(t_1^2+t_2^2+t_3^2+t_4^2)^2-4(t_1^2t_3^2+t_2^2t_4^2+2t_1t_2t_3t_4)=0.\hspace{-5pt} 
\end{align}
The possible solutions for the above equation are either $t_1=t_3,\;t_2=t_4$ or $t_1=-t_3,\;t_2=-t_4$. The other condition (b) leads to equation  
\begin{align}
t_1^2t_2^2+t_2^2t_4^2-2t_1t_2t_3t_4=0\Rightarrow \frac{t_1t_3}{t_2t_4}=1.
\end{align}
Thus, the two topological transitions of the Hermitian model are obtained at $t_4=t_2,t^2/t_2$ for the choice of parameters $t_1=t_3=t$. The topological phases in the Hermitian quadripartite model are characterized by the Zak phase \cite{zak_berrys_1989}, which is again quantized at either $0$ or $\pi$ due to the inversion symmetry of the model \cite{xiao_surface_2014}. We show the Zak phase values by dashed black lines in Fig.~\ref{fig:phase1} and Fig.~\ref{fig:15}, where the value of $2\pi$ is equivalent to $0$ when the module $2\pi$ is applied.
\subsection{\label{cEP}Exceptional points}
Each degeneracy at band touching of any two bands in the Hermitian model splits into two EPs in the non-Hermitian model ($t_{4r}\neq t_{4l}$). The EPs in the non-Hermitian systems are calculated from the non-Hermitian degeneracy conditions: (a) $E_1(\pi)=E_2(\pi)$ or $E_3(\pi)=E_4(\pi)$ and (b) $E_2(0)=E_3(0)$.
The first condition results into a relation:
\begin{align}
(t_{1l}t_{1r}+t_{2l}t_{2r}+t_{3l}t_{3r}+t_{4l}t_{4r})^2-4(t_{1l}t_{1r}t_{3l}t_{3r}+\nonumber\\t_{2l}t_{2r}t_{4l}t_{4r}+t_{1r}t_{2r}t_{3r}t_{4r}+t_{1l}t_{2l}t_{3l}t_{4l})=0.
\end{align}
For the particular choice of parameters, $t_{1l}=t_{1r}=t_{3r}=t_{3l}=t$ and $t_{2l}=t_{2r}=t_2$ and $t_{4l}=t_4e^{\theta}$, $t_{4r}=t_4$, the details of two solutions ($t_{m_{1}},t_{m_2}$) for $t_4$ of the above equation are given in Eqs.~\ref{eq:16},\ref{eq:17}. The second condition  leads to the equation
 \begin{align}
 (t_{1l}t_{1r}t_{3l}t_{3r}+t_{2l}t_{2r}t_{4l}t_{4r}-t_{1r}t_{2r}t_{3r}t_{4r}-t_{1l}t_{2l}t_{3l}t_{4l})=0.
 \end{align}
 The two solutions for the above equation denoting two more EPs are 
 \begin{align}
 t_{1r}t_{3r}=t_{2l}t_{4l},\; t_{1l}t_{3l}=t_{2r}t_{4r},
 \end{align}
which can be written for $t_{4}$ within our above choice of parameters as $t_{z_{1,2}}
=\frac{t^2}{t_2},\frac{t^2e^{-\theta}}{t_2}$.
\subsection{\label{cDispersion}Dispersive energy spectra}
\begin{figure}[h!]
\includegraphics[width=0.4925\linewidth]{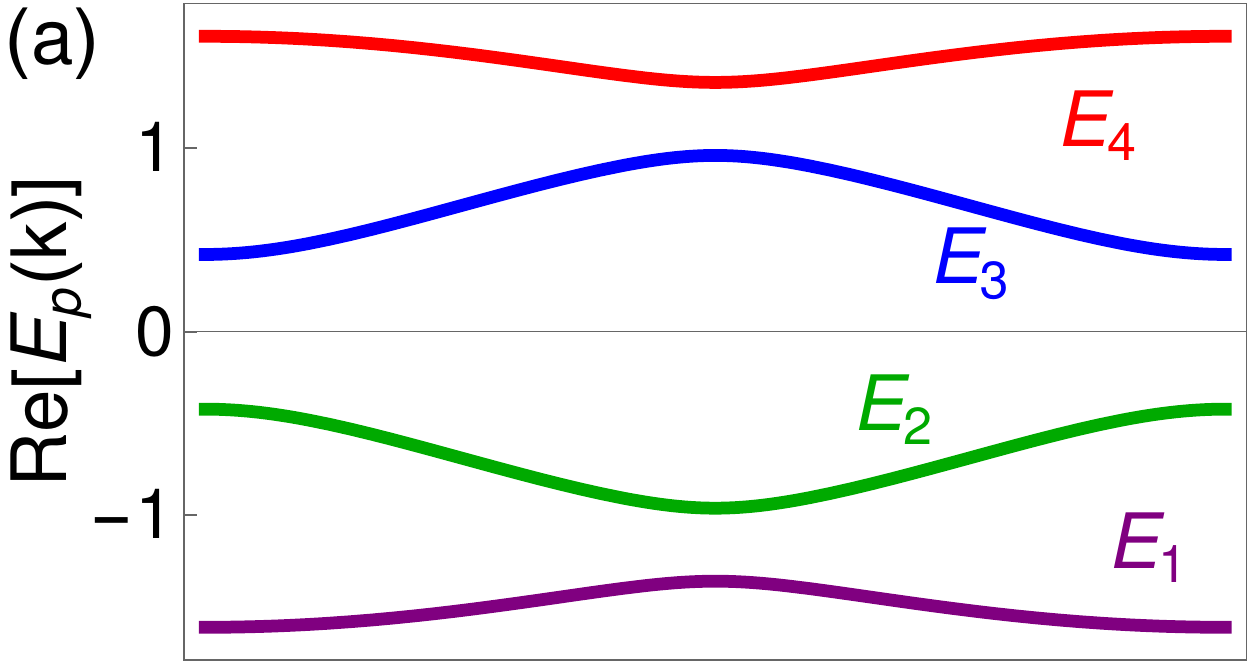}
\includegraphics[width=0.4925\linewidth]{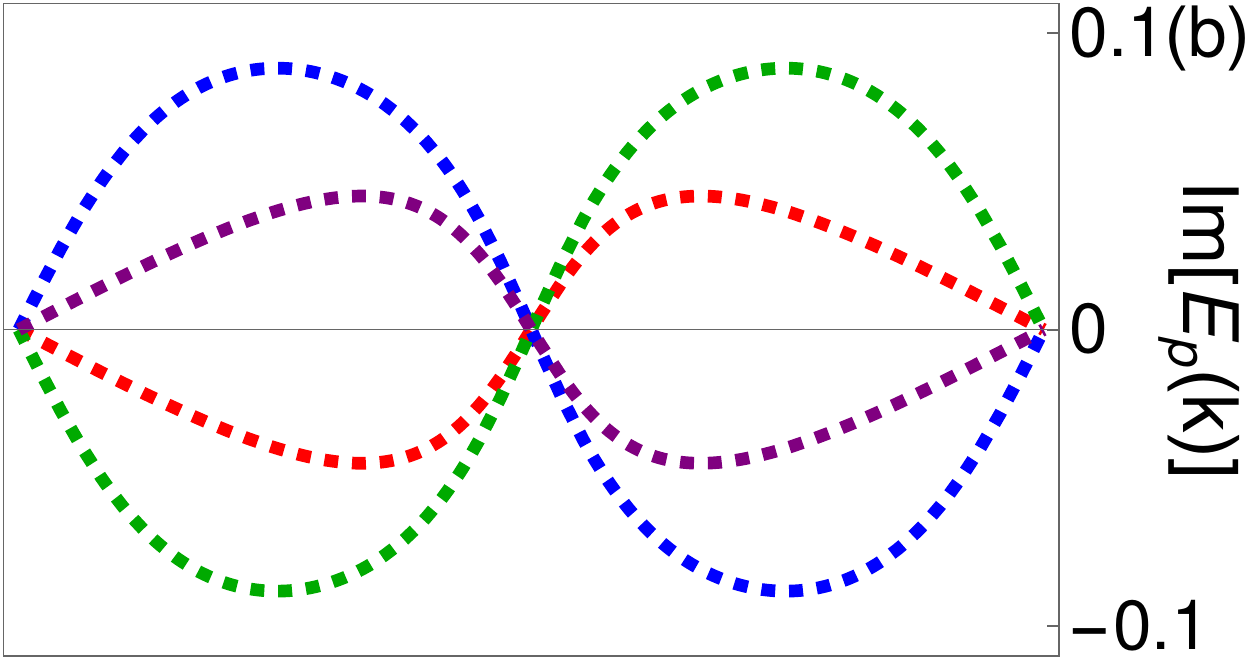}
\includegraphics[width=0.4925\linewidth]{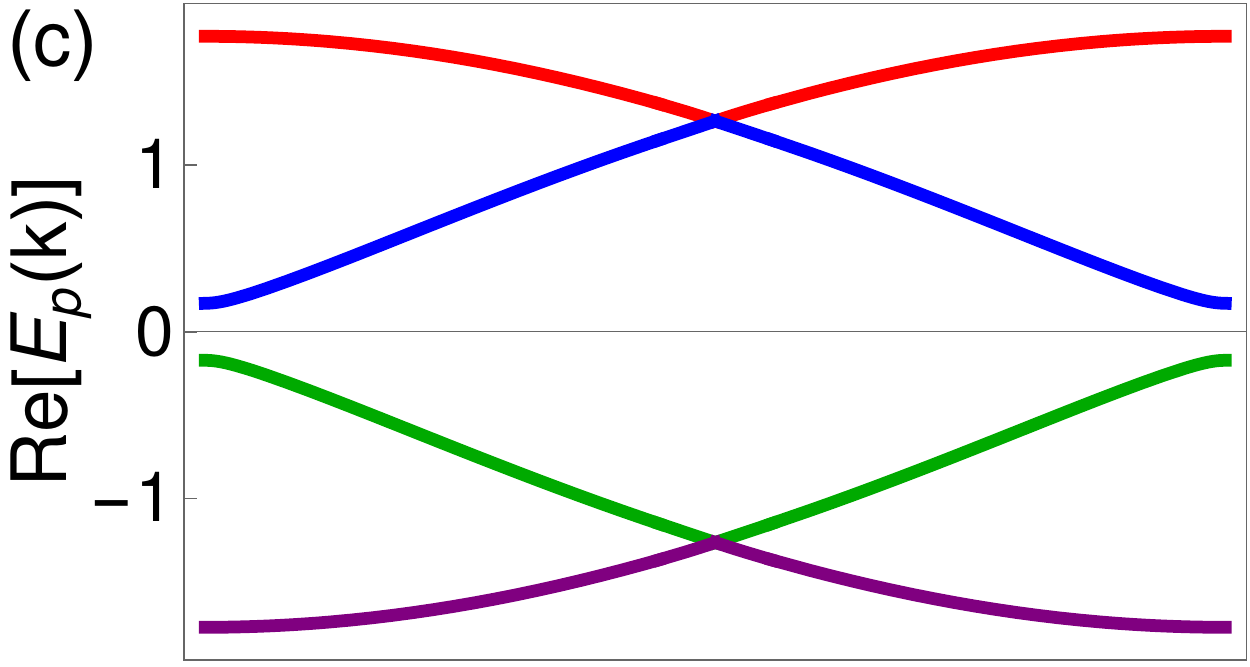}
\includegraphics[width=0.4925\linewidth]{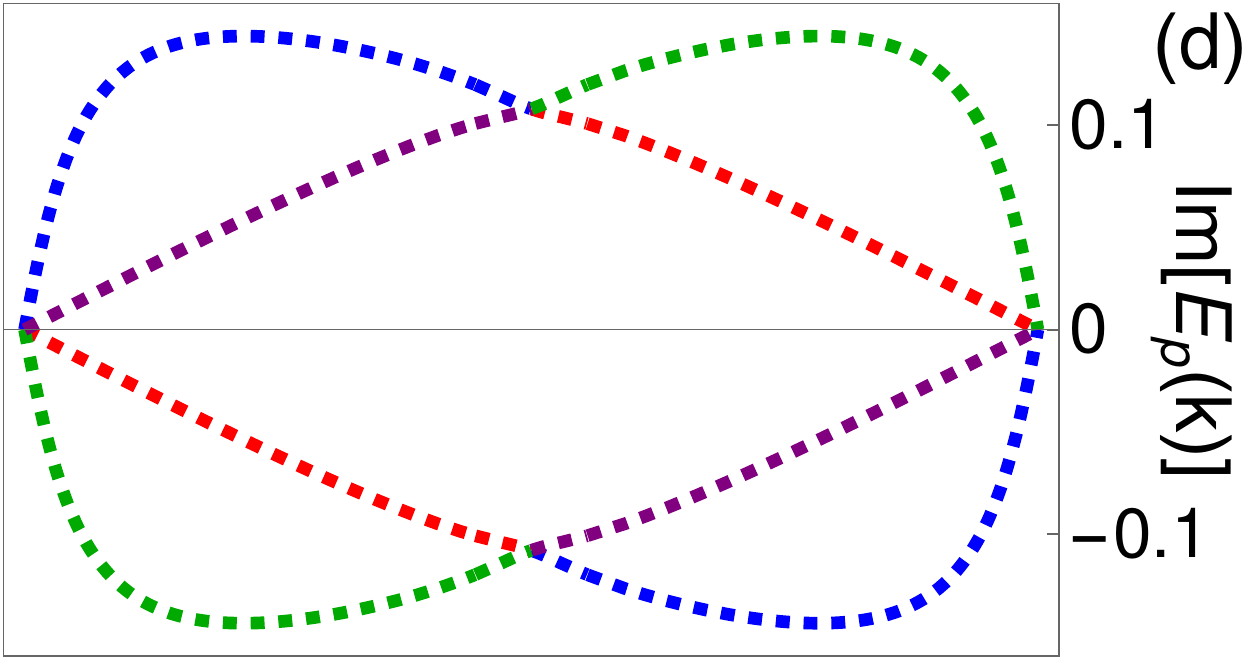}
\includegraphics[width=0.4925\linewidth]{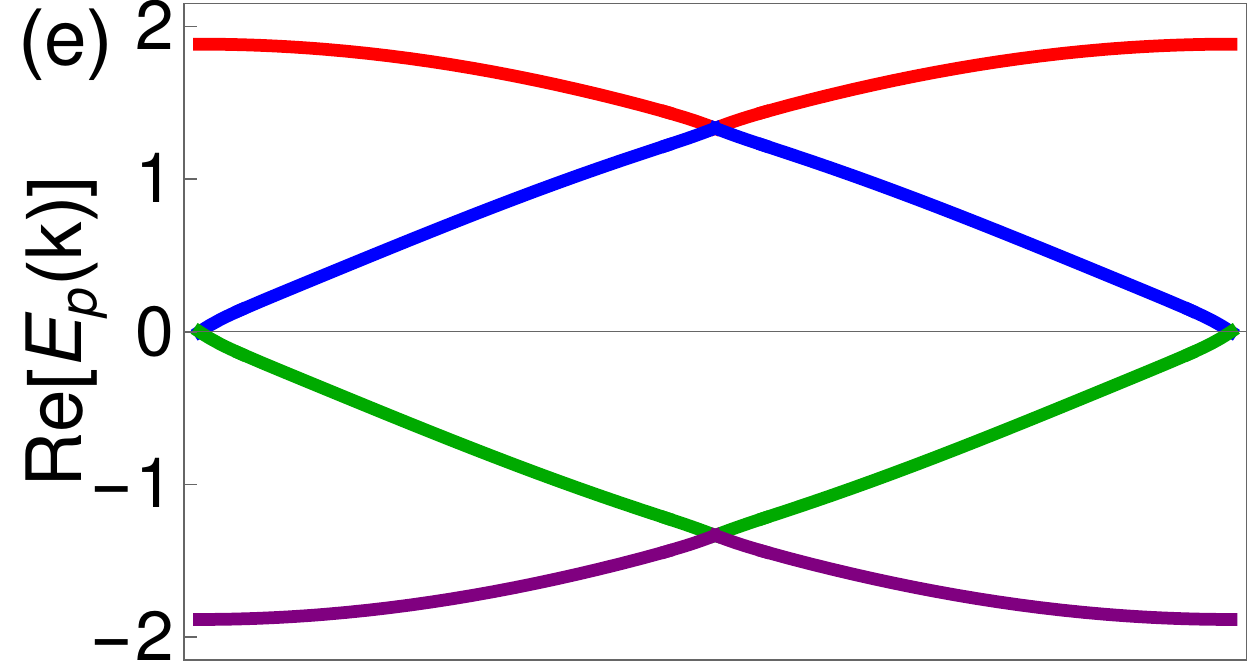}
\includegraphics[width=0.4925\linewidth]{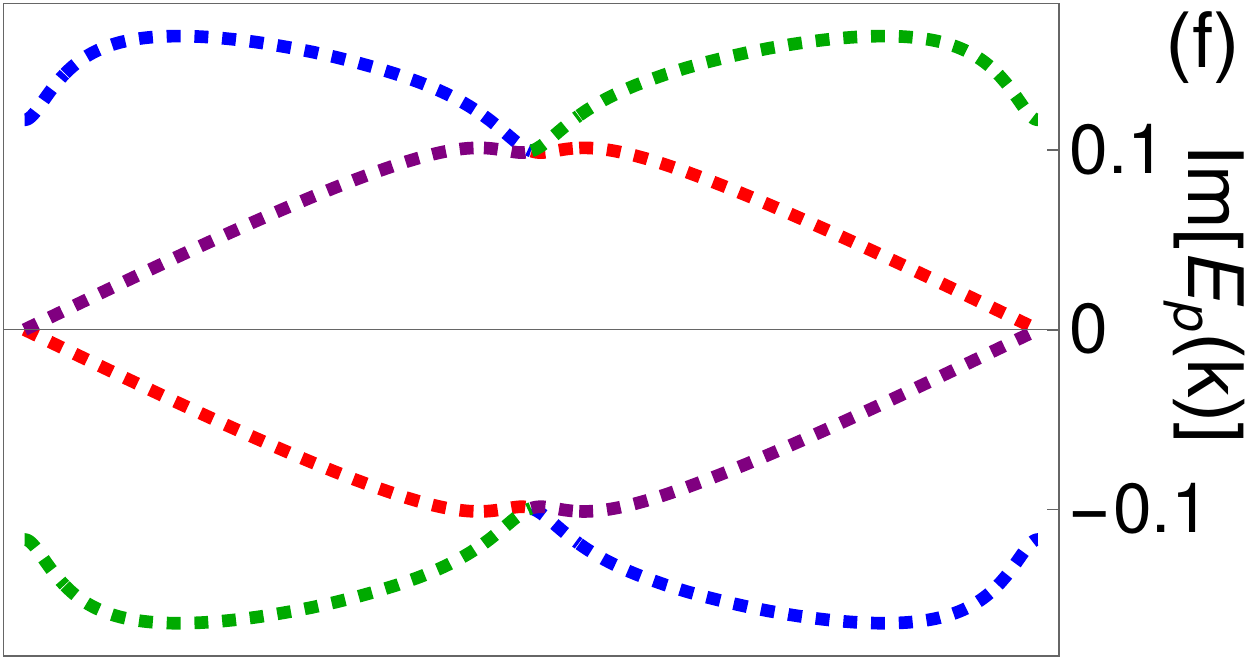}
\includegraphics[width=0.4925\linewidth]{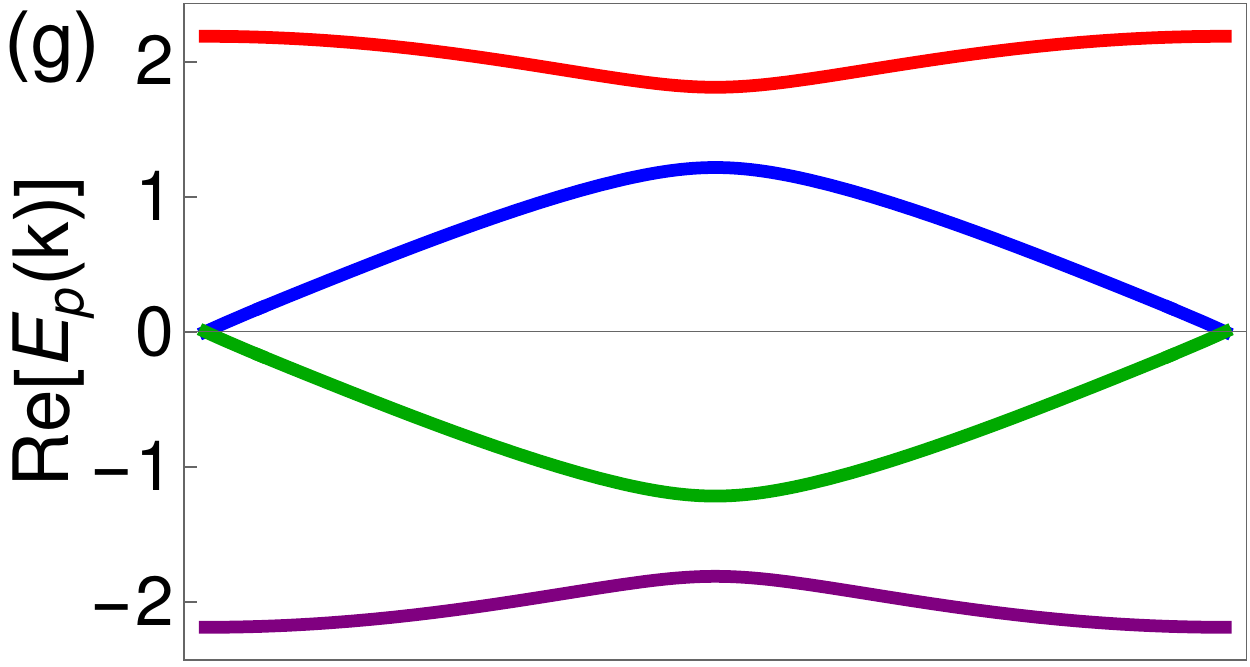}
\includegraphics[width=0.4925\linewidth]{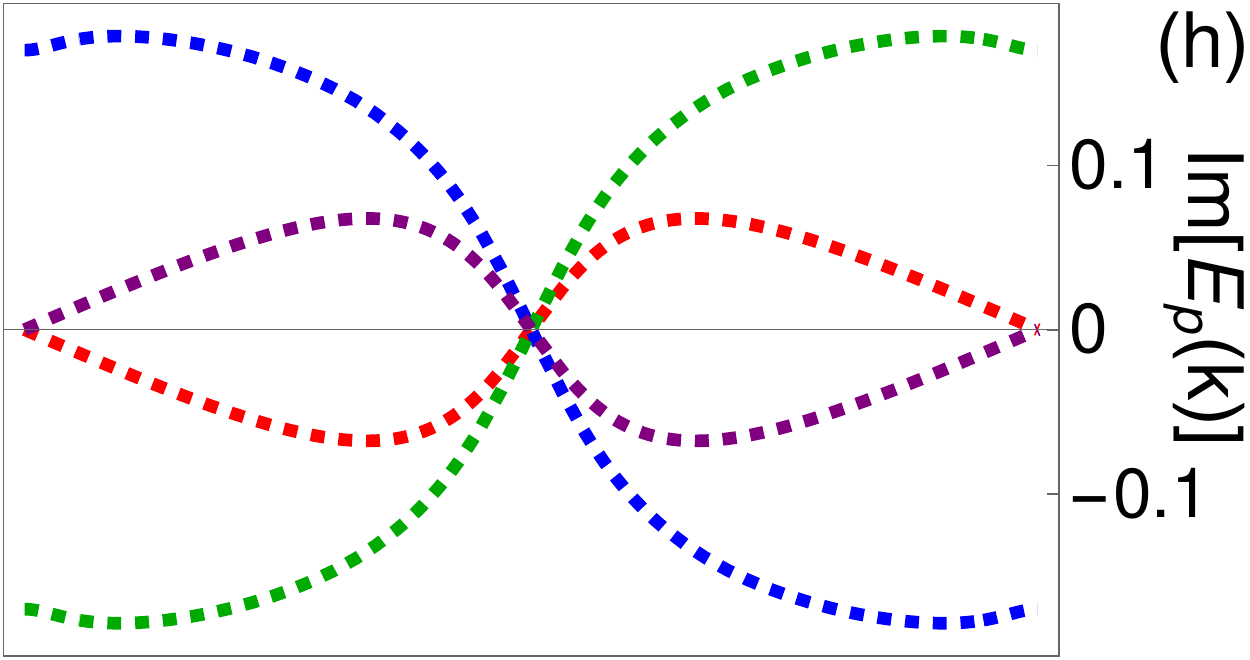}
\includegraphics[width=0.4925\linewidth]{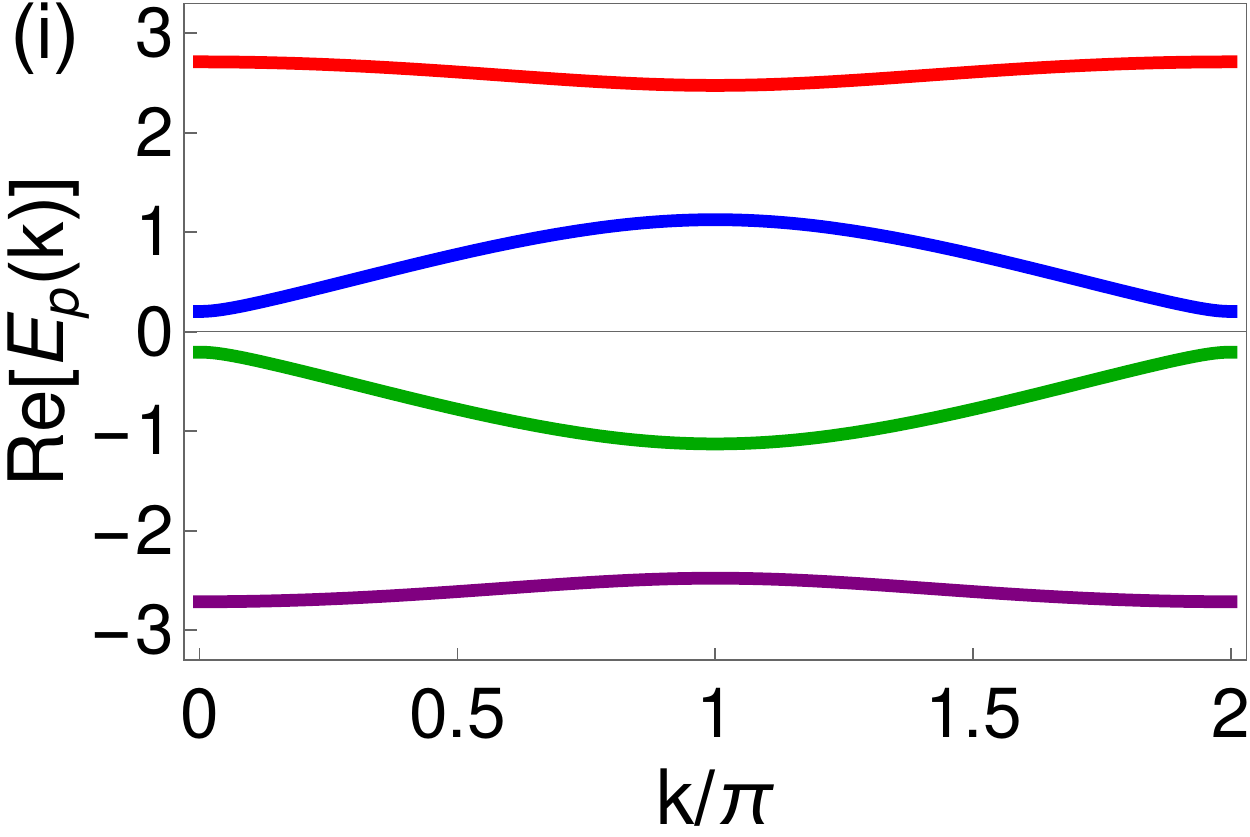}
\includegraphics[width=0.4925\linewidth]{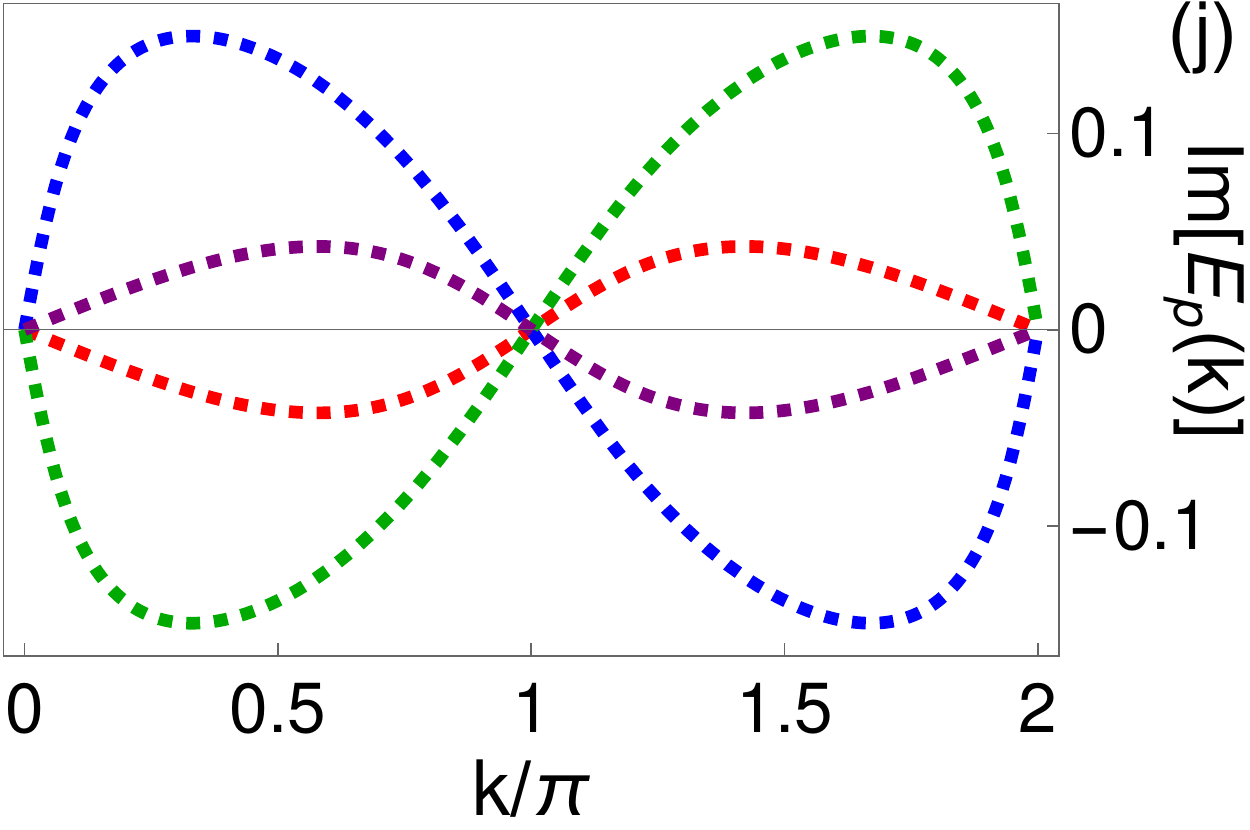}
\caption{The dispersive energy spectra of the nSSH4 model. The four energy bands $E_1,\;E_2,\;E_3,\;E_4$ are denoted by purple, green, blue, red lines, respectively. The solid (dotted) line depicts the real (imaginary) part of the  spectrum. The parameters are $t=1.0, t_{2}=0.8, \theta=0.75$ in all plots and (a,b) $t_4=0.25$ ($T$), (c,d) $t_4=0.5$ ($CI_m$), (e,f) $t_4=0.65$ ($CM_3$), (g,h) $t_4=1.0$ ($CI_z$) and (i,j) $t_4=1.5$ ($NT$).}
\label{fig:disp}
\end{figure}

The presence of above mentioned EPs are confirmed by the dispersive energy spectra  of the nSSH4 model using Eq.~\ref{eq:15}  in Fig.~\ref{fig:disp}.  
The band closing in real energy and the discontinuity in the imaginary energy of the bands at different $k$ points validate the encircling of the EPs in the parametric space shown in Fig.~\ref{fig:intro}. 
The individual bands forming separate loops in Figs.~\ref{fig:intro}(a,e) are shown in Figs.~\ref{fig:disp}(a,b) and Figs.~\ref{fig:disp}(i,j), which represent the topologically trivial ($T$) and nontrivial ($NT$) insulating phases, respectively. Two composite loops centered at finite energies in Fig.~\ref{fig:intro}(b) are formed by dispersive energy bands in Figs.~\ref{fig:disp}(c,d), showing band touchings at real energies
 $\pm\sqrt{(t_{1r}t_{1l}+t_{1r}t_{1l}+t_{2r}t_{2l}+t_{3r}t_{3l}+t_{4r}t_{4l})/2}$. The composite loop encircling all the EPs in Fig.~\ref{fig:intro}(c) is formed by  all the dispersive energy bands in Figs.~\ref{fig:disp}(e,f) showing the band touchings at different $k$ values. The composite loop centered at zero energy in Fig.~\ref{fig:intro}(d) is created by the dispersive energy bands in Figs.~\ref{fig:disp}(g,h), which shows only the central two bands touching at zero real energy. 
 \subsection{\label{cHP}Hybrid points}
The other interesting transition points in the non-Hermitian systems are hybrid points~\citep{shen_topological_2018}. In contrast to the EPs and the degeneracy points, these HPs are not developed due to the band touching. However, the Hamiltonian is defective at these points like the EPs. Therefore, the HPs for nSSH4 are calculated using the normalization constant of the bi-orthogonal eigenvectors given in Eq.~\ref{eq:19}, i.e., $N_p(k)=2E_{p}^2(k)^2(2E^2_{p}(k)-t_{1l}t_{1r}-t_{2l}t_{2r}-t_{3l}t_{3r}-t_{4l}t_{4r})(E_p^{2}(k)-t_{1l}t_{1r}-t_{2l}t_{2r})=0$. The HPs are derived by condition, $E_p(k)=\pm\sqrt{t_{1l}t_{1r}+t_{2l}t_{2r}}$. It is observed that this condition on energies is satisfied at $k=\pi$ for different bands such that $E_{1,4}(k=\pi)=\mp\sqrt{t_{1l}t_{1r}+t_{2l}t_{2r}}$ gives a solution $t_{4l}=\frac{t_{2r}t_{3r}}{t_{1l}}$ and $E_{2,3}(k=\pi)=\mp\sqrt{t_{1l}t_{1r}+t_{2l}t_{2r}}$ gives a solution $t_{4r}=\frac{t_{2l}t_{3l}}{t_{1r}}$. These HPs for our particular choice of parameters are given at $\tilde{t}_{m_{1}}=t_2e^{-\theta}$ for bands $E_{1,4}$ and $ \tilde{t}_{m_{2}}={t_2}$ for bands $E_{2,3}$.
\subsection{\label{ctopo}Topological phases}
\begin{figure}[h!]
\centering
(a)\hspace{-15pt}\includegraphics[scale=0.67]{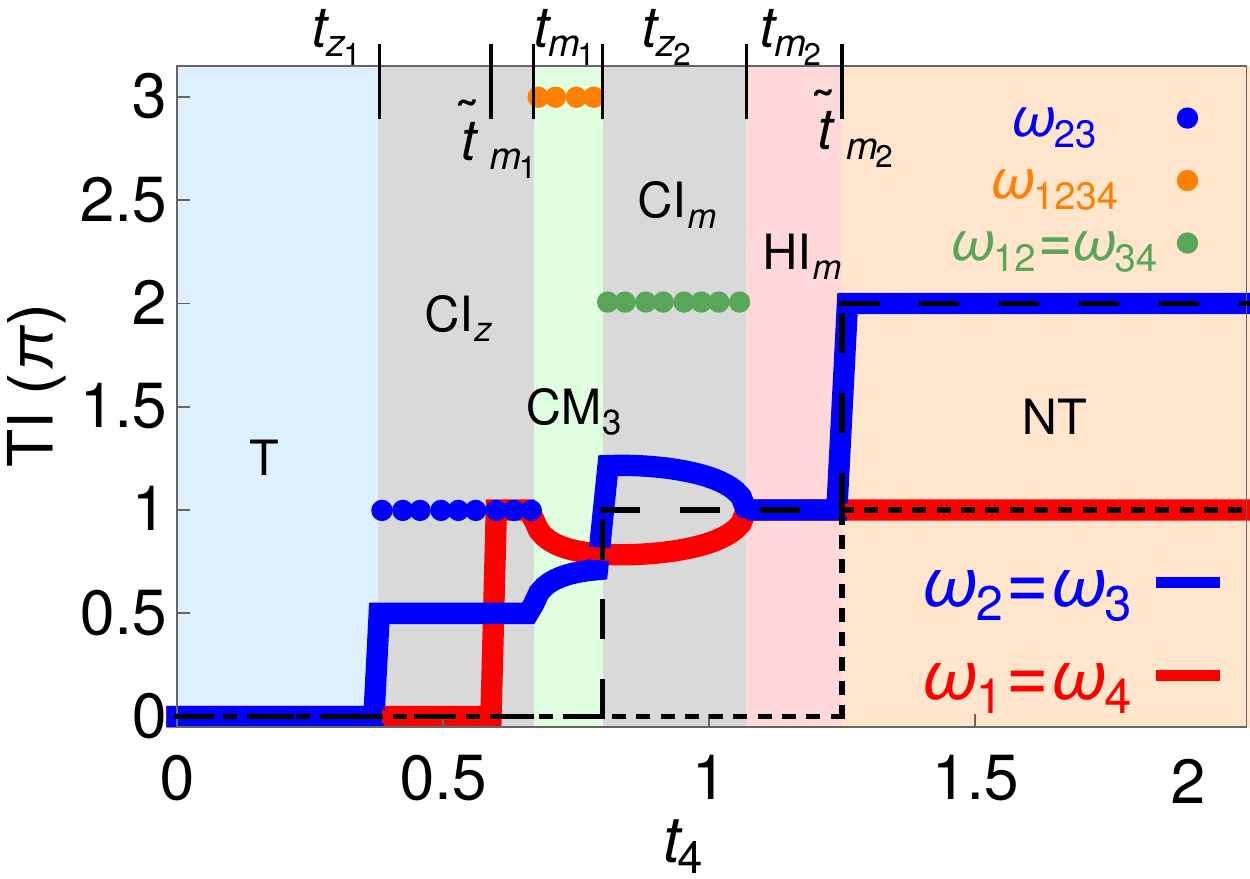}
(b)\hspace{-15pt}\includegraphics[scale=0.67]{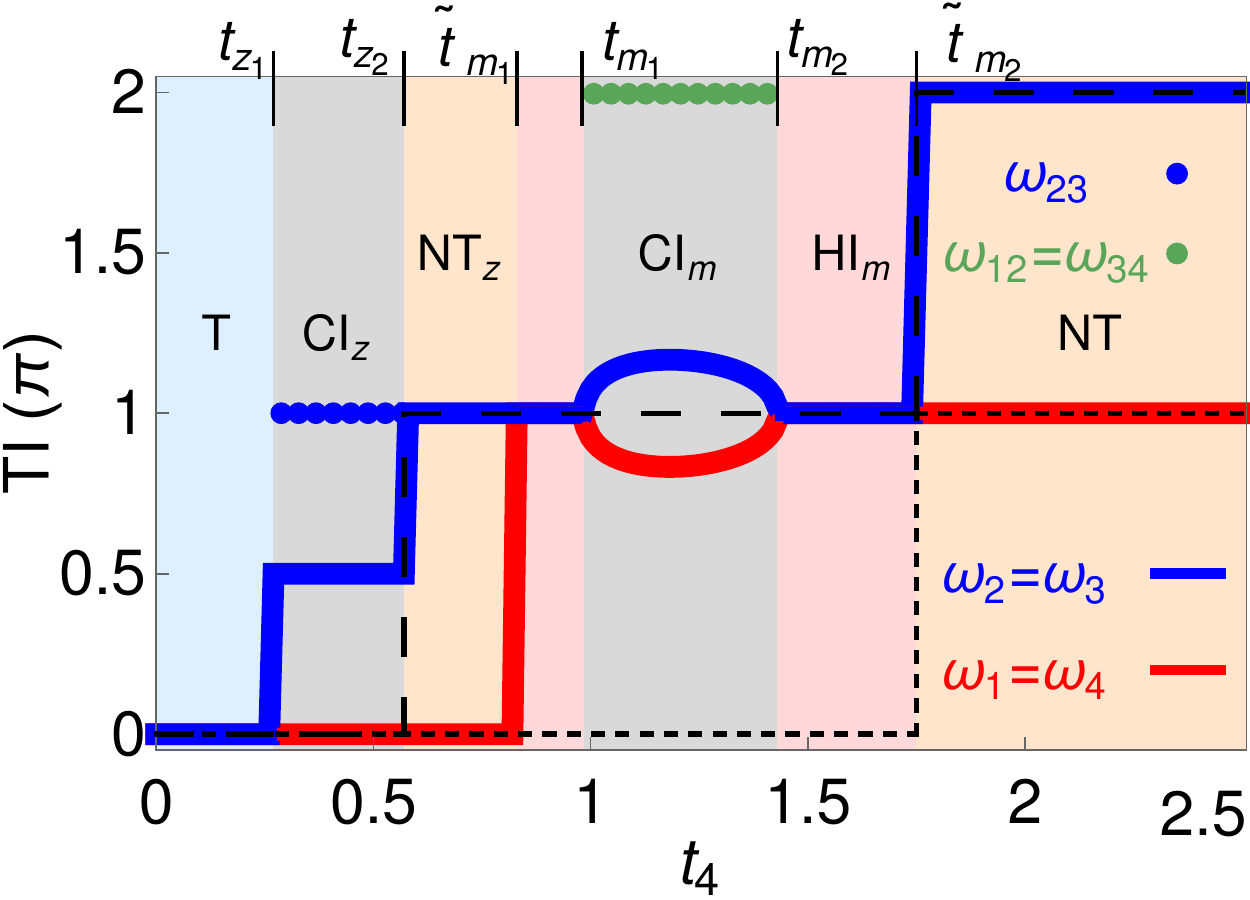}
\caption{Topological invariant (TI), $\omega_p, \omega_{p_1p_2\dots p_n}$ for the complex energy bands of the nSSH4 model with inter-cell hopping $t_4$. The parameters are $t=1.0, \theta=0.75$ in all plots and (a) $t_{2}=1.25$ and (b) $t_{2}=1.75$. We show the topologically trivial ($T$) phase by light blue, composite insulating ($CI_z, CI_m$) phases by light grey, composite metallic ($CM_3$) phase by light green, hybrid insulating ($HI_m$) phase by light red, and topologically nontrivial ($NT$) phase by light orange. The short and long dashed lines indicate the Zak phase in
the Hermitian limit $(\theta =0)$ for $p= 1,4$ and $p = 2,3$, respec-
tively. The dotted lines represent the composite geometric phases in the composite insulating and metallic phases.}
\label{fig:15}
\end{figure}

We now discuss the topological phases of the nSSH4 model for the following set of parameters $t_{1r}=t_{1l}=t_1=t,\;t_{3r}=t_{3l}=t_3=t$, $t_{2r}=t_{2l}=t_2>\sqrt{t_1t_3}$ with $t_{4r}=t_4$ and $t_{4l}=t_4e^{\theta}$.
The values of the relevant topological invariants for the above set of parameters are shown in Fig.~\ref{fig:15}.
The complex Zak phase $\omega_p$ for all the individual bands is zero in the trivial ($T$) insulating phase for $t_4<t_{{z_1}}$, and is quantized at $\pi$ ($E_{1,4}$) and $2\pi$ ($E_{2,3}$) in the nontrivial ($NT$) insulating phase for $t_4>\tilde{t}_{m_2}$ in Figs.~\ref{fig:15}(a,b). 
The composite geometric (Zak) phase $\omega_{p_1p_2\hdots p_n}$ is used as a topological invariant for different composite insulating and metallic phases. In Fig.~\ref{fig:15}(b), $\omega_{23}=\pi$ for $t_{4}\in(t_{z_1},t_{z_2})$ indicates a topologically nontrivial composite insulating ($CI_z$) phase, and $\omega_{12}=\omega_{34}=2\pi$ for $t_{4}\in(t_{m_1},t_{m_2})$ predicts a topologically trivial composite insulating ($CI_m$) phase. These composite phases in Fig.~\ref{fig:15}(b ) are separated by a mixed nontrivial insulating ($NT_z$) phase with $\omega_{1,4}=0,\omega_{2,3}=\pi$. However, this $NT_z$ phase disappears when the two composite phases overlap for a specific choice of parameters shown in Fig.~\ref{fig:15}(a). This leads to the creation of a composite loop formed by all the energy bands together shown in Fig.~\ref{fig:intro}(c), and the loop represents a composite metallic ($CM_3$) phase encircling three EPs. A non-zero quantized value of $\omega_{1234}=3\pi~(\tilde{\omega}_{1234}=\pi)$ indicates nontrivial topology of $CM_3$ for $t_4\in(t_{m_1},t_{z_2})$ in Fig.~\ref{fig:15}(a). The overlap of $CI_z$ and $CI_m$  redefines the boundaries of the other composite phases in Fig.~\ref{fig:15}(a). 

Apart from the $T$, $NT$, composite phases, the non-Hermiticity gives rise to hybrid insulating ($HI_m$) phase between the HPs and the EPs. Due to the lack of degeneracies at the HPs, the complex energy bands are represented by individual separate loops for $t_{4}\in(\tilde{t}_{m_1},t_{m_1})$ and $t_{4}\in(t_{m_2},\tilde{t}_{m_2})$. Thus, the topology of these $HI_m$ phases is studied by the complex Zak phase, which indicates a quantized and non-zero value of $\pi$ for all the bands in Fig.~\ref{fig:15}(b). Therefore, the $HI_m$ phases are topologically nontrivial. The topology of the system should be explored applying the composite geometric (Zak) phase for any overlap of the $CI_z$ phase with these HP boundaries. The formation of the composite loops is indicated by $\omega_{23}=\pi$ for $t_4\in(\tilde{t}_{m_1},t_{m_1})$ in Fig.~\ref{fig:15}(a) and for $t_4\in(t_{m_2},\tilde{t}_{m_2})$ in Fig.~\ref{fig:phase1}.
 
  The full list of topological phases, their boundaries and the relevant topological invariants in different phases are summed up in Table~\ref{tabel}.
\begin{table}[h!]
 \begin{center}
      \begin{tabular}{|c|c|c|}
     \hline
 \textbf{No.} &\textbf{Phases} & \textbf{TIs($t_2>t$)} \\
 \hline 
 \multirow{2}{*}{$1.$}  &\textbf{Trivial ($T$)}&  \\
  &$t_4\in(0,t_{z_1})$&$\omega_p=0$, $p\in All$\\
   \hline
 \multirow{1}{*}{2a.}  &\textbf{Composite Insulator($CI_z$)}&  \\
 &$t_4\in( t_{z_1},t_{z_2})$ &$\omega_{1,4}=0,\;\omega_{23}=\pi$ \\
 \cline{2-3}
 {2b.} &\textbf{Composite Insulator($CI_z$)}&  \\
 &$t_4\in( t_{z_1},\tilde{t}_{m_1})$ &$\omega_{1,4}=0,\;\omega_{23}=\pi$ \\
  \hline
  \multirow{1}{*}{3a.} &\textbf{Nontrivial($NT_z$) }&  \\
  &$t_4\in(t_{z_2},\tilde{t}_{m_1})$ & $\omega_{1,4}=0$, $\omega_{2,3}=\pi$  \\\cline{2-3}
  {3b.} &\textbf{Composite Insulator($CI_z$)}&  \\
  &$t_4\in(\tilde{t}_{m_1},t_{m_1})$ &$\omega_{1,4}=\pi$, $\omega_{23}=\pi$  \\
  \hline
    \multirow{1}{*}{4a.}  &\textbf{Hybrid Insulator($HI_m$)}&  \\
  &$t_4\in(\tilde{t}_{m_1},t_{m_1})$ &$\omega_{p}=\pi$, $p\in All$  \\
  \cline{2-3}
{4b.} &\textbf{Composite Metallic($CM_3$)}&  \\
 &$t_4\in(t_{m_1},t_{z_2})$ &$\omega_{1234}=3\pi$\\
 \hline
 \multirow{1}{*}{5a.}  &\textbf{Composite Insulator($CI_m$)}&  \\
 &$t_4\in(t_{m_1},t_{m_2})$ &$\omega_{12}=2\pi$, $\omega_{34}=2\pi$\\\cline{2-3}
 {5b.}&\textbf{Composite Insulator($CI_m$)}&  \\
 &$t_4\in(t_{z_2},t_{m_2})$ &$\omega_{12}=2\pi$, $\omega_{34}=2\pi$\\
 \hline
 \multirow{2}{*}{6.}  &\textbf{Hybrid Insulator($HI_m$)}&  \\
  &$t_4\in(t_{m_2},\tilde{t}_{m_2})$ &$\omega_{p}=\pi$, $p\in All$ \\
  \hline
 \multirow{2}{*}{7.}  &\textbf{Nontrivial($NT$)}&  \\
  &$t_4\in(\tilde{t}_{m_2},\infty)$ &$\omega_{1,4}=\pi$, $\omega_{2,3}=2\pi$ \\
  \hline
     \end{tabular}
  \end{center}
\setlength{\abovecaptionskip}{-2pt}  
  \caption{The topological phases, their boundaries and the relevant topological invariants of nSSH4 model for parameters $t_{1l}=t_{1r}=t_{3l}=t_{3r}=t$, $t_{2l}=t_{2r}=t_2>t$, $t_{4l}e^{-\theta}=t_{4r}=t_4$, and (a) without and (b) with mixing of two types of composite insulating phases. Here, $\omega_{p_1p_2\hdots p_s}$ (Eq. 7) and  $\omega_{p_1,p_2}=\omega_{p_1}=\omega_{p_2}$ (Eq.~\ref{eq:5}) are composite and complex geometric (Zak) phases, respectively.}
  \label{tabel}
\end{table}

A similar summary of various topological phases of the nSSH4 model for the parameters $t_{1r}=t_{1l}=t_1=t,\;t_{3r}=t_{3l}=t_3=t$, $t_{2r}=t_{2l}=t_2<\sqrt{t_1t_3}$ with $t_{4r}=t_4$ and $t_{4l}=t_4e^{\theta}$ is presented in Table~\ref{tabel1}. 
\begin{table}[h!]
 \begin{center}
      \begin{tabular}{|c|c|c|}
     \hline
 \textbf{No.} &\textbf{Phases} & \textbf{TIs($t_2<t$)} \\
 \hline 
 \multirow{2}{*}{$1.$}  &\textbf{Trivial ($T$)}&  \\
  &$t_4\in(0,\tilde{t}_{m_1})$&$\omega_p=0$, $p\in All$\\
   \hline
 \multirow{2}{*}{2.}  &\textbf{Hybrid Insulator ($HI_{m}$)}&  \\
 &$t_4\in( \tilde{t}_{m_1},t_{m_1})$ &$\omega_{1,4}=\pi,\;\omega_{2,3}=0$ \\
  \hline
  \multirow{1}{*}{3a.} &\textbf{Composite Insulator($CI_m$) }&  \\
  &$t_4\in(t_{m_1},t_{m_2})$ & $\omega_{12}=\pi$, $\omega_{34}=\pi$  \\\cline{2-3}
  {3b.} &\textbf{Composite Insulator($CI_m$)}&  \\
  &$t_4\in(t_{m_1},t_{z_1})$ &$\omega_{12}=\pi$, $\omega_{34}=\pi$  \\
  \hline
    \multirow{1}{*}{4a.}  &\textbf{Hybrid Insulator ($HI_{m}$)}&  \\
  &$t_4\in(t_{m_2},\tilde{t}_{m_2})$ &$\omega_{1,4}=\pi$, $\omega_{2,3}=0$  \\
  \cline{2-3}
 {4b.}&\textbf{Composite Metallic($CM_3$)}&  \\
 &$t_4\in(t_{z_1},t_{m_2})$ &$\omega_{1234}=3\pi$\\
 \hline
 \multirow{1}{*}{5a.}  &\textbf{Nontrivial($NT_m$)}&  \\
 &$t_4\in(\tilde{t}_{m_2},t_{z_1})$ &$\omega_{p}=\pi$, $p\in All$\\\cline{2-3}
 {5b.}&\textbf{Composite Insulator($CI_z$)}&  \\
 &$t_4\in(t_{m_2},\tilde{t}_{m_2})$ &$\omega_{1,4}=\pi$, $\omega_{23}=\pi$\\
 \hline
 \multirow{1}{*}{6a.}  &\textbf{Composite Insulator($CI_z$)}&  \\
 &$t_4\in(t_{z_1},t_{z_2})$ &$\omega_{1,4}=\pi$, $\omega_{23}=3\pi$\\\cline{2-3}
 {6b.}&\textbf{Composite Insulator($CI_z$)}&  \\
 &$t_4\in(\tilde{t}_{m_2},t_{z_2})$ &$\omega_{1,4}=\pi$, $\omega_{23}=3\pi$\\
 \hline
 \multirow{2}{*}{7.}  &\textbf{Nontrivial($NT$)}&  \\
  &$t_4\in(t_{z_2},\infty)$ &$\omega_{1,4}=\pi$, $\omega_{2,3}=2\pi$ \\
  \hline
     \end{tabular}
  \end{center}
  \setlength{\abovecaptionskip}{-2pt}
  \caption{The topological phases, their boundaries and the relevant topological invariants of nSSH4 model for parameters $t_{1l}=t_{1r}=t_{3l}=t_{3r}=t$, $t_{2l}=t_{2r}=t_2<t$, $t_{4l}e^{-\theta}=t_{4r}=t_4$, and (a) without and (b) with mixing of two types of composite insulating phases.}
  \label{tabel1}
\end{table} 

The composite phases $CI_z$ and $CM_3$ are topologically nontrivial in Tables~\ref{tabel},~\ref{tabel1}, and the respective composite loops enclose one and three EPs. But, the composite insulating phase $CI_m$, represented by a composite loop enclosing one EP, is topologically trivial with $\omega_{12}=\omega_{34}=2\pi~(\tilde{\omega}_{12}=\tilde{\omega}_{34}=0)$ when the $CI_m$ phase appears after the $CI_z$ phase for $t_2>t$, which is in sharp contrast to the case $t_2<t$ shown in Fig.~\ref{fig:phase1}.
\vspace{-10pt}
\subsection{\label{spenrose}Geometric realizations of composite band-structures}
 In this part, we give geometric realizations of the band-structures of composite insulating and metallic states through some popular geometric objects. The topology of two-band composite loops with $4\pi$ periodicity on the parametric space can be realized by a simple M{\"o}bius strip made with a piece of paper~\citep{vyas_topological_2021}. 
 \begin{figure}[h!]
\includegraphics[width=0.9\linewidth, height=0.35\linewidth]{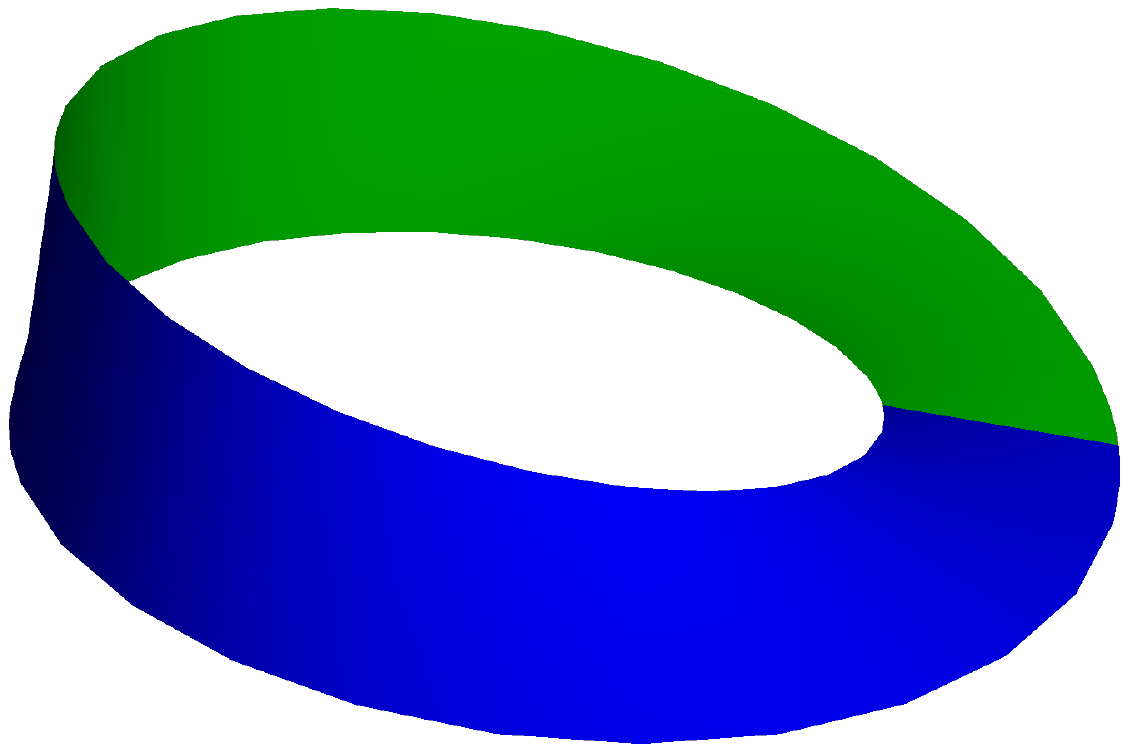}
\setlength{\belowcaptionskip}{-12pt}
\caption{The M{\"o}bius strip as a geometric realization of band-structure of composite state $CI_z$ of the non-Hermitian quadripartite SSH model. The green and blue colors  indicate two sides of the paper which are glued together after a half-twist ($\pi$ rotation).}
\label{mobius}
 \end{figure}
 The M{\"o}bius strip related to the composite loop in Fig.~\ref{fig:intro}(d) for the composite phase $CI_z$ is shown in Fig.~\ref{mobius}. Let's take a piece of paper with two different colored sides (front and back), and join the two ends of the paper to form a loop after giving a half-twist, i.e., $\pi$ rotation at one end of the paper. A line drawn along the edge of the strip returns to its starting point after traveling a double length of the original strip. Therefore, the M{\"o}bius strip and the composite loop in $CI_z$ phase possess a periodicity of $4\pi$. The change in color after one rotation on M{\"o}bius strip represents an exceptional point. The same analogy also works for the other composite loops denoting $CI_m$ phase.
 
\begin{figure}[h!]
{(a)\includegraphics[width=0.445\linewidth]{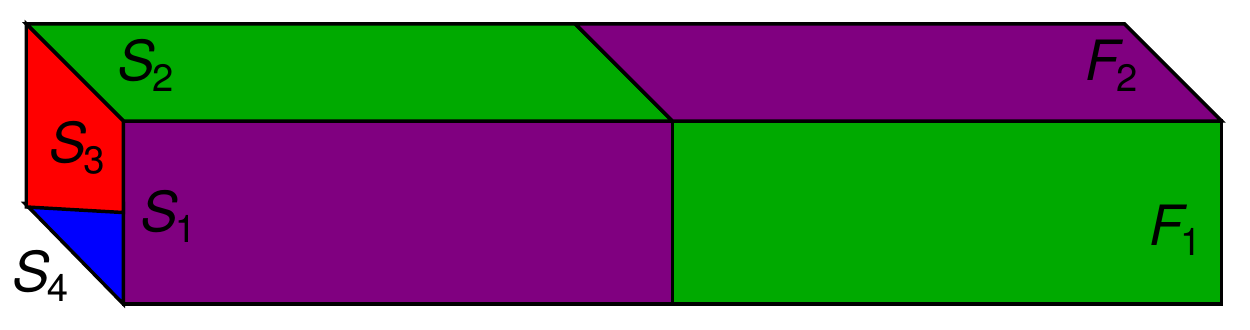}
 (b)\includegraphics[width=0.445\linewidth]{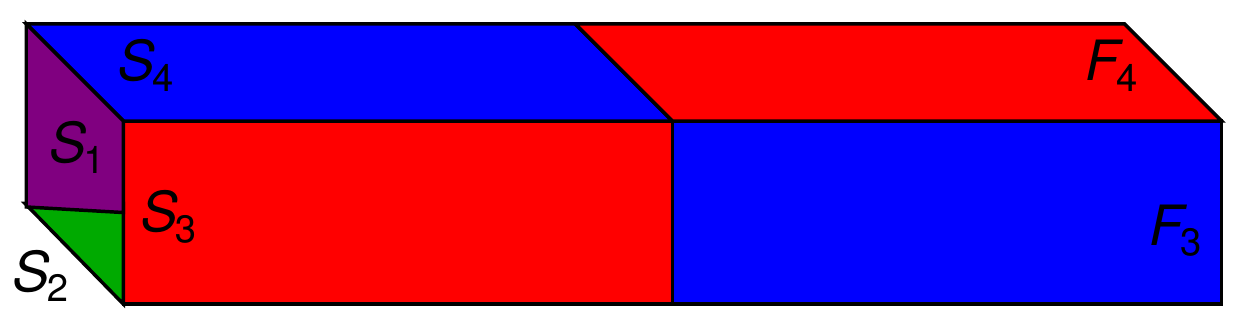}}\\
 {(c)\includegraphics[width=0.44\linewidth]{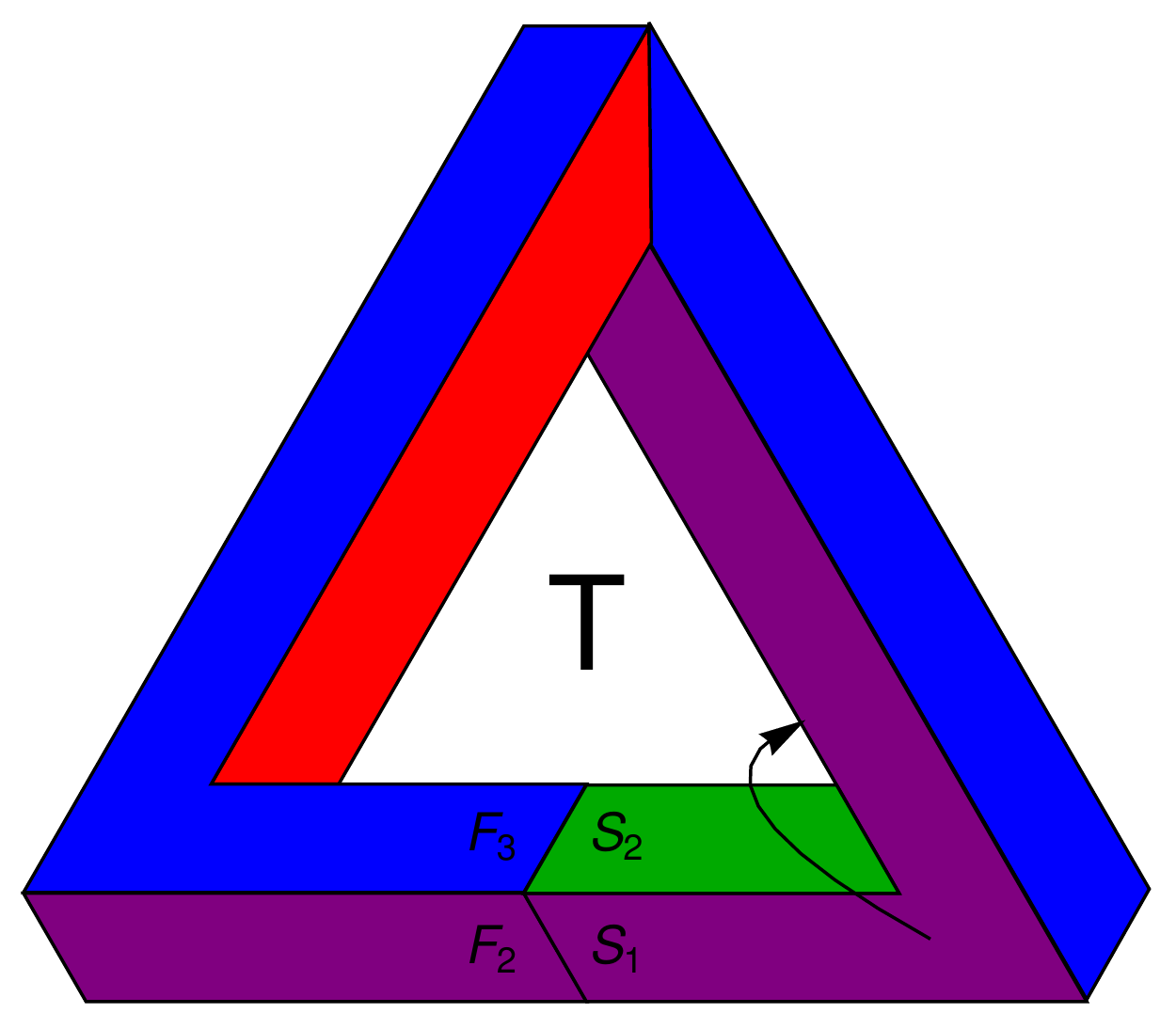}
 (d)\includegraphics[width=0.45\linewidth]{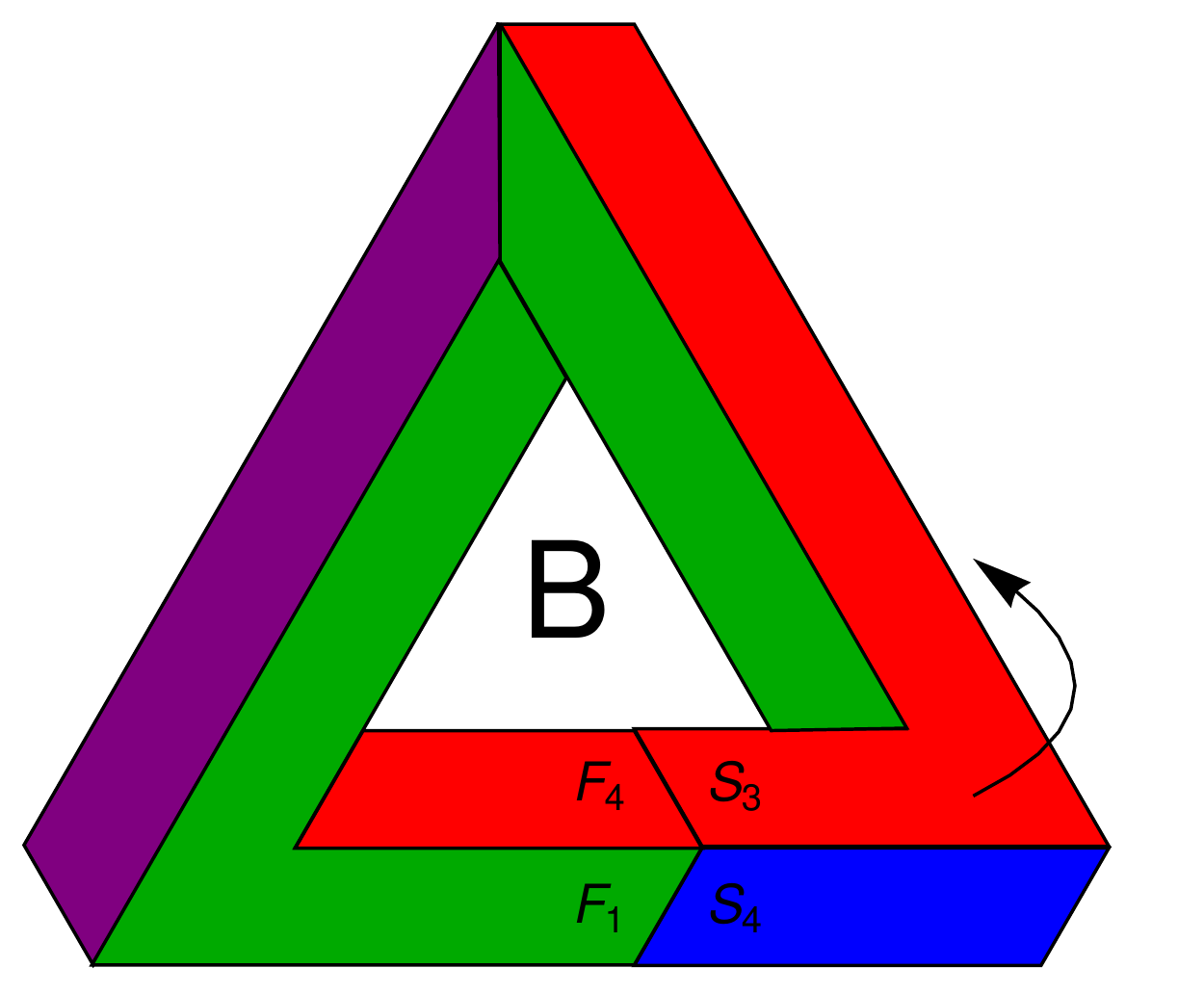}}
 \setlength{\belowcaptionskip}{-10pt}
 \caption{Geometric realization of the band-structure of composite metallic state $CM_3$ of the quadripartite non-Hermitian SSH model. (a, b) Four sides of the bar  whose each side is painted with two colors. ($S_1,S_2,S_3,S_4$) and ($F_1,F_2,F_3,F_4$) denote planes at two ends. (c,d) Top and bottom view of the Penrose triangle formed by joining the two ends of the bar after giving three quarter twists.}
\label{penrose} 
 \end{figure}

 Similarly, the topology of composite metallic $CM_3$ state (Fig.~\ref{fig:intro}c) can be realized by another popular geometric object, the Penrose triangle, shown in Fig.~\ref{penrose}.   
 Let's consider a four-sided flexible bar whose each side is painted with two colors, e.g., violet-green, green-violet, red-blue, blue-red as shown in Figs.~\ref{penrose}(a,b). The planes at the two ends of the bar are named as ($S_1,S_2,S_3,S_4$) and ($F_1,F_2,F_3,F_4$). The two ends of the bar is joined together after giving three quarter twists (each twist is $\pi/2$ rotation with respect to the axis of the bar) at one end of the bar. Each corner of the Penrose triangle in Figs.~\ref{penrose}(c,d) represents a one-quarter twist to the bar. The change in color on the same plane indicates an exceptional point. A line drawn on a side of the Penrose triangle returns to its starting point after traveling a quadruple length of the original bar. Therefore, the Penrose triangle and the composite loop in $CM_3$ phase have a periodicity of $8\pi$.
  \vspace{-5pt}
\subsection{\label{cfourth}Fourth-order exceptional point}
We find that the nSSH4 model features a fourth-order EP when $\alpha^2-\beta|_{k=\pi}=0$, $\beta|_{k=\pi}=0$ in Eq.~\ref{eq:15}.
\begin{figure}[h!]
\includegraphics[width=0.65\linewidth]{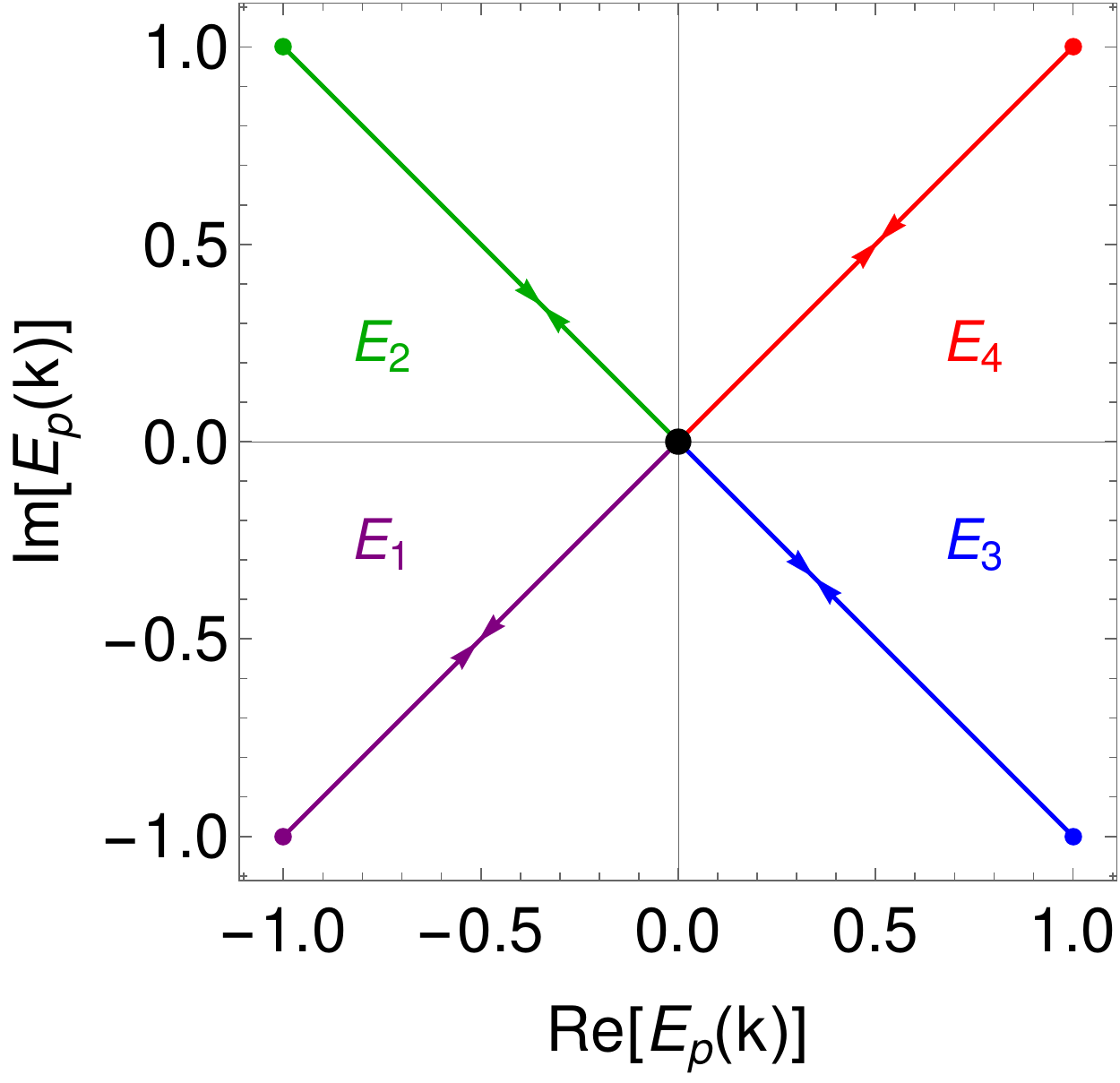}
\caption{The complex energy bands of the nSSH4 model on the parametric space for $t_{1l}=t_{3r}=t$, $t_{1r}=t_{3l}=-t$, $t_{2l}=t_{4r}=te^{-\frac{\theta}{2}}$ and $t_{2r}=t_{4l}=te^{\frac{\theta}{2}}$. The origin (black dot) depicts the fourth-order exceptional point.}
\label{fig:para}
\end{figure}
 Then, all four energy bands touch each other at $E_p(k=\pi)=0$. The above two conditions for the fourth-order EP can be achieved by one such choice of parameters:  $t_{1l}=t_{3r}=t$, $t_{1r}=t_{3l}=-t$, $t_{2l}=t_{4r}=te^{-\frac{\theta}{2}}$ and $t_{2r}=t_{4l}=te^{\frac{\theta}{2}}$, which also lacks the intrinsic inversion symmetry employed in the earlier choice of parameters in  our study of second-order EPs in the quadripartite model. 
The complex band energies of nSSH4 model on the parametric space around the fourth-order EP are shown in Fig.~\ref{fig:para}, which appears different from the previous parametric plots with second-order EPs.

\subsection{\label{all_boundary}Parametric enegry plots for different boundary conditions}
\begin{figure}[h!]
\centering
\includegraphics[width=\linewidth, height=1.25\linewidth]{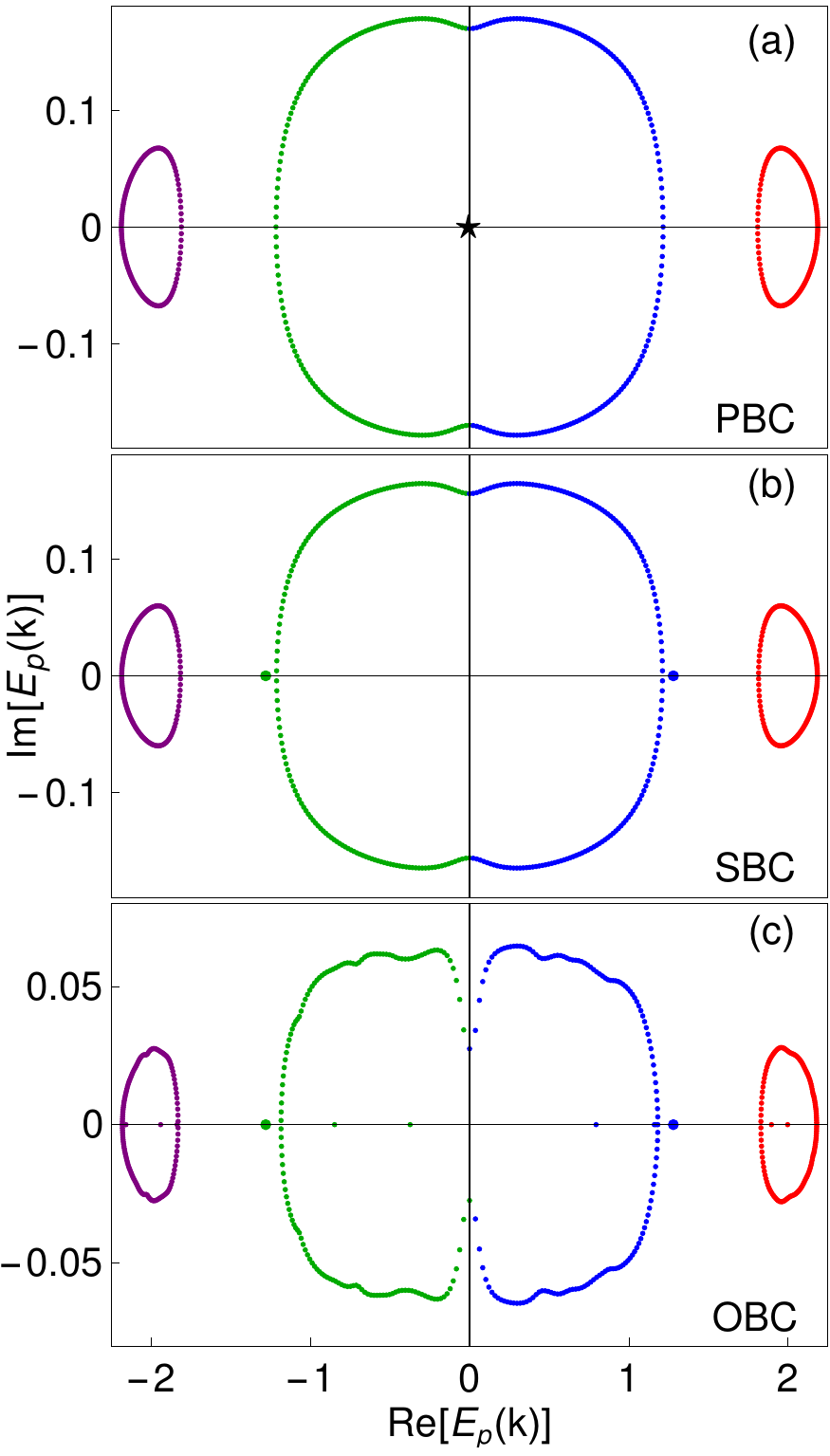}     
\caption{The complex energy spectra of nSSH4 model on the parametric space of Re[$E_p(k)$] and Im[$E_p(k)$] for $L=150$, $t=1.0$, $t_2=0.8$, $t_4=0.65$ and $\theta=0.75$ under (a) periodic ($\eta=1$),  (b)  special ($\eta=1/L$) and (c) open ($\eta=0$) boundary conditions. The energy of four bands $E_1,\;E_2,\;E_3,\;E_4$ are depicted by purple, green, blue, red colors, respectively. The big dots show the energies of edge states.} 
\label{fig:allboundary}
\end{figure}

The boundaries of non-Hermitian systems play a vital role in restoring conventional bulk-boundary correspondence for energy spectra calculated using periodic and open boundary conditions~\citep{jin_bulk-boundary_2019,ghatak_new_2019,xiao_coexistence_2017,yao_edge_2018,delplace_zak_2011,xiong_why_2018}. 
The absence of the chiral-inversion symmetry in such systems leads to the non-Hermitian Aharonov-Bohm (AB) effect~\citep{jin_bulk-boundary_2019,yao_edge_2018} under periodic boundary conditions (PBC) and various kinds of skin effects~\citep{ghatak_new_2019,okugawa_non-hermitian_2021,zhang_correspondence_2020,bergholtz_exceptional_2021} under open boundary conditions (OBC). The bulk boundary correspondence in this system can be restored using  special boundary conditions (SBC)~\citep{vyas_topological_2021,PhysRevLett.127.116801}, unlike the generalized Brillouin zone method~\citep{yao_edge_2018,jin_bulk-boundary_2019,zhu_mathcalpt_2014,kunst_biorthogonal_2018}.
To keep the consistency with Fig.~\ref{fig:intro}(d), we here show energy spectra of composite insulating ($CI_z$) phase under PBC ($\eta=1$), SBC ($\eta=1/L$) and OBC ($\eta=0$) in Fig.~\ref{fig:allboundary}.  
The composite loops calculated with PBC and SBC are very similar to each other in Figs.~\ref{fig:allboundary}(a,b). The big dots denote the edge states in the topologically non-trivial phase for SBC and OBC. 
 Nevertheless, the parametric plot with OBC for the composite loop in Fig.~\ref{fig:allboundary}(c) is a bit different from them in Figs.~\ref{fig:allboundary}(a,b).

\section{\label{symmetry}Symmetries}
The study of topology of a model is inconclusive without understanding its symmetries. We here discuss the symmetries of multipartite non-Hermitian SSH chains. In our present study, we have particularly applied intrinsic inversion symmetry of unit cells \citep{zak_berrys_1989,xiao_coexistence_2017} for quantized topological invariants (bi-orthonormal geometric phase for separate or composite loops) and sublattice symmetry for real values of the topological invariants. The interplay of non-Hermiticity with topology gives rise 38-fold non-Hermitian topological classes~\citep{gong_topological_2018,kawabata_symmetry_2019,kawabata_topological_2019} replacing usual 10-fold Altland-Zirnbauer (AZ) symmetry classes~\citep{altland_nonstandard_1997,chiu_classification_2016,ryu_topological_2010}. The unification and ramification of different symmetries in various topological systems due to the lack of Hermiticity generate these new topological classes~\citep{kawabata_symmetry_2019,kawabata_topological_2019}.  The non-Hermitian SSH chain is one of the simplest example for the symmetry unification and ramification discussed in Ref.~\citep{kawabata_symmetry_2019}. The Hamiltonian of  the multipartite non-Hermitian SSH model can be represented in a generic off-diagonal form along with doubling of the space (for nSSH3) or rearrangement of the basis (for nSSH4): 
\begin{align}
\mathcal{H}_s(k)=\begin{bmatrix}
\mathbf{O} &\mathbf{h}^s_1(k)\\
\mathbf{h}^s_2(k) &\mathbf{O}
\end{bmatrix},
\end{align} 
where $\mathbf{O}$ and $\mathbf{h}^s_{1,2}$ are matrices of dimension 1, 3, 2 for $s=2,3,4$ sites per unit cell, respectively. Here, $\mathbf{O}$ is a null matrix, and the matrix $\mathbf{h}^s_{1,2}$ for different cases are given by
\begin{align}
\mathbf{h}^2_1=t_{1l}+t_{2r}e^{ik},\;\mathbf{h}^2_2=t_{1r}+t_{2l}e^{-ik};
\end{align}
\begin{align}
\mathbf{h}^3_1=\begin{bmatrix}
0 &t_{1l} &t_{3r}e^{2ik}\\
t_{3l} &t_{2r}&0\\
t_{1r} &0 &t_{2l}
\end{bmatrix},\;\mathbf{h}^3_2=\begin{bmatrix}
0 &t_{3r} &t_{1l}\\
t_{1r} &t_{2l}&0\\
t_{3l}e^{-2ik} &0 &t_{2r}
\end{bmatrix};
\end{align}
\vspace{-20pt}
\begin{align}
\mathbf{h}^4_1=\begin{bmatrix}
t_{1l} &t_{4r}e^{ik}\\
t_{2r} &t_{3l}
\end{bmatrix},\;\mathbf{h}^4_2=\begin{bmatrix}
t_{1r} &t_{2l}\\
t_{4l}e^{-ik} &t_{3r}
\end{bmatrix}.
\end{align}
For the Hamiltonian of our system with real parameters ($t_{\sigma l}, t_{\sigma r}$), we have $\mathcal{H}_s^*(-k)=\mathcal{H}_s(k)$, which implies time-reversal symmetry (TRS) in our system since $\mathcal{T}_+\mathcal{H}_s(k)\mathcal{T}^{-1}_+=\mathcal{H}_s(-k)$, for an anti-unitary operator $\mathcal{T}_+=K$. The TRS imposes constraint on the energy spectra of the system such that the energies appear as a pair of $E,\;-E^*$. Another symmetry in our system is sublattice symmetry (SLS), which follows the relation $\mathcal{S}\mathcal{H}_s(k)\mathcal{S}^{-1}=-\mathcal{H}_s(k)$. The unitary operator for SLS is given by $\mathcal{S}=\sigma_z\otimes\mathcal{I}$, where  $\mathcal{I}$ represents the identity matrix with dimension 1, 3, 2 for $s=2,3,4$ sites per unit cell, respectively. 
The other important symmetry of the system is particle-hole symmetry (PHS$^\dagger$), which is described by an anti-unitary operator $\mathcal{T}_-=K\mathcal{S}$ such that $\mathcal{T}_-\mathcal{H}_s(k)\mathcal{T}^{-1}_-=-\mathcal{H}_s(-k)$.

Compatible with the unification of TRS and PHS$^\dagger$ and the ramification of chiral  symmetry (CS) and SLS, the non-Hermitian SSH models fall in the equivalent classes of AI and D$^{\dagger}$ both with $\mathcal{S}_+$~\citep{kawabata_symmetry_2019,kawabata_topological_2019}. Here, $\mathcal{S}_+$ represents the class with SLS commutating with TRS and PHS$^\dagger$, i.e., $[\mathcal{T_\pm,S}]=0$. Along with these symmetry classifications, our system indicates a real line gap ($L_r$) shown in the parametric plots of the complex energy spectra, which further classify the system on the basis of complex energy gaps~\citep{kawabata_parity-time-symmetric_2018,kawabata_symmetry_2019}. The topology of such classes is characterized by a topological invariant belonging to class $\mathbb{Z}$ (integer values)~\citep{ cayssol_topological_2021}. 

\section{\label{vorticity}Vorticity}
\begin{figure}[h!]
(a)\hspace{-15pt}\includegraphics[scale=0.34]{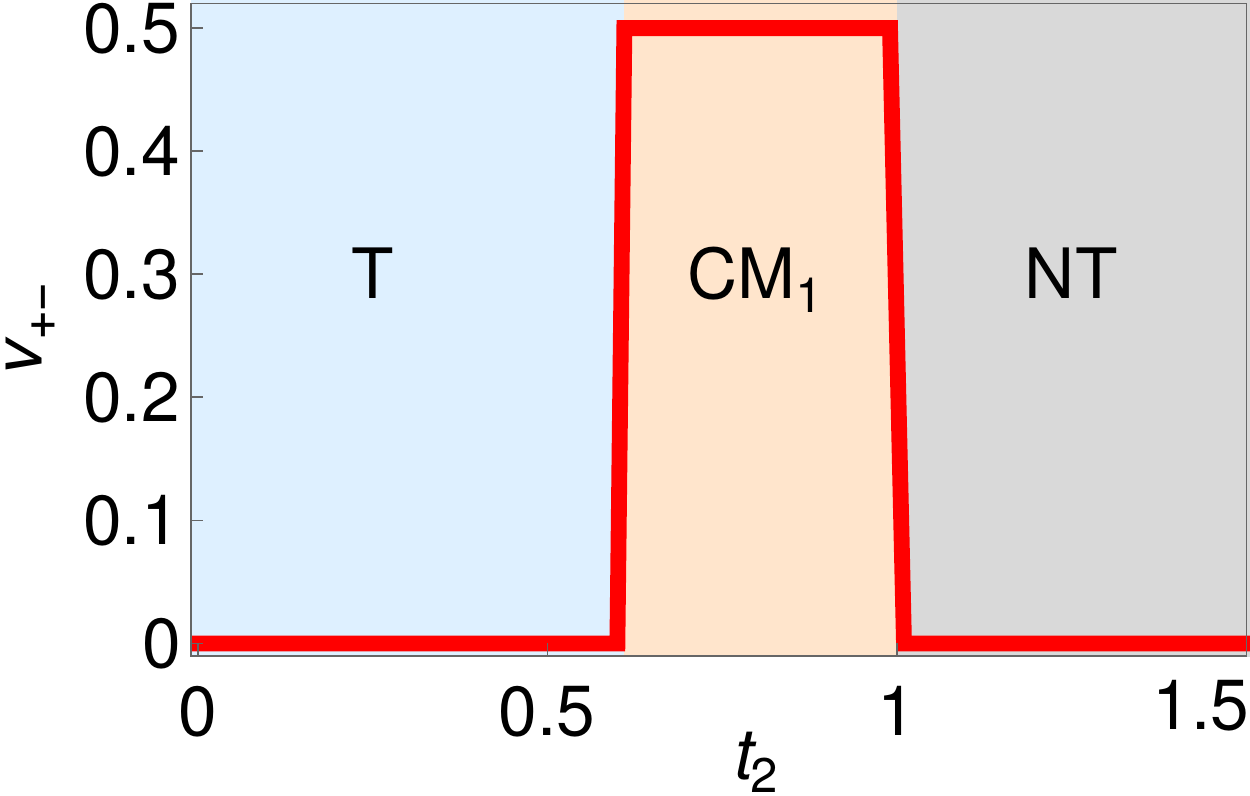}
(b)\hspace{-10pt}\includegraphics[scale=0.335]{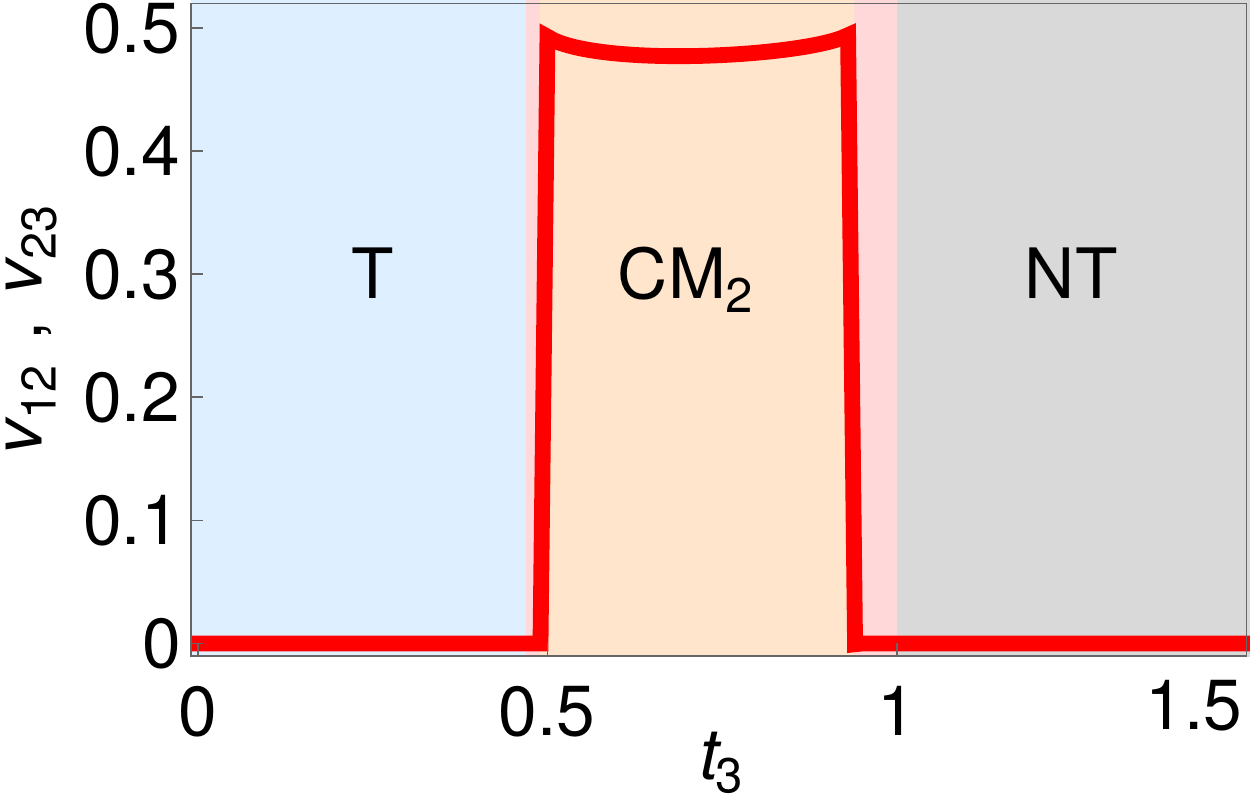}
\caption{(a) The vorticity ($\nu_{+-}$) for nSSH2 model as a function of $t_2$ and the other parameters are $t_{1l}=t_{1r}=1$, $t_{2l}=t_2e^{\theta}$, $t_{2r}=t_2$ and $\theta=0.5$. (b) The vorticity ($\nu_{12},\;\nu_{23}$) for nSSH3 model as a function of $t_3$ and the other parameters are $t_{1l}=t_{1r}=t_{2r}=t_{2l}=t=1$, $t_{3l}=t_3e^{\theta}$, $t_{3r}=t_3$ and $\theta=0.75$. }
\label{fig:vorticity}
\end{figure}

Here, we discuss a new topological invariant coined as vorticity~\citep{shen_topological_2018,ghatak_new_2019,ghatak_observation_2020,
yin_geometrical_2018}, which is uniquely defined for the non-Hermitian energy bands, unlike various geometric phases calculated using cell-periodic Bloch eigenfunctions of the system. The vorticity for a pair of energy bands is defined as 
\begin{align}
\nu_{ij}=-\frac{1}{2\pi}\displaystyle\int_{0}^{2\pi}\triangledown_k\;\text{arg}[E_i(k)-E_j(k)]\;dk,
\end{align} 
where $i,j$ are band indices.  
A half-integer value of the vorticity over a closed contour indicates the underlying band degeneracy due to the presence of EPs. The Hamiltonian of a system is defective if its eigenfunctions are non-analytic or singular. The topological phase transition points on the parametric space can be classified into four categories depending on the value of vorticity and the analytic properties of the  eigenfunctions of the Hamiltonian in a non-Hermitian model~\citep{shen_topological_2018}. These are: 1. EP (when vorticity is half-integer and the Hamiltonian is defective), 2. HP (when vorticity is zero and the Hamiltonian is defective), 3. Dirac point (when vorticity is zero and the Hamiltonian is non-defective), 4. Vortex point (when vorticity is non-zero integer and the Hamiltonian is non-defective). We have applied the above nomenclatures in our description of the boundaries of different composite metallic and insulating phases in the nSSH models. The vorticity for the nSSH2 and nSSH3 models are depicted in Fig.~\ref{fig:vorticity}.
We observe that the vorticity for the two bands of the nSSH2 model (Fig.~\ref{fig:vorticity}(a)) traces the full composite metallic phase ($CM_1$) with a value of 0.5 over the whole phase, which is similar to the bi-orthonormal Zak phase of the individual energy bands in Fig.~\ref{fig:5}. The boundaries with a vorticity of half-integer indicate the appearance of EPs. Therefore, the boundary of $CM_1$ is formed by the EPs. Furthermore, the vorticity of the nSSH3 model captures the boundaries of the phase $CM_2$ for both pairs of bands, i.e., $E_1(k),\;E_2(k)$ and $E_2(k),\;E_3(k)$, in Fig.~\ref{fig:vorticity}(b). The boundaries of the hybrid insulating phases in the nSSH3 and nSSH4 models are created by HPs as the Hamiltonian is defective at these boundaries and the vorticity is zero. 
\section{\label{braid}Braids and knots}
The braiding of complex energy bands is another important topological characteristics of the non-Hermitian systems~\citep{wang_topological_2021}. The complex energy spectra of non-Hermitian systems form braid trajectories along momentum over the first Brillouin zone.
\begin{figure}[h!]
(a)\includegraphics[scale=0.75]{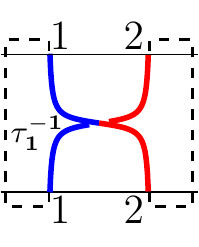}
(b)\includegraphics[scale=0.75]{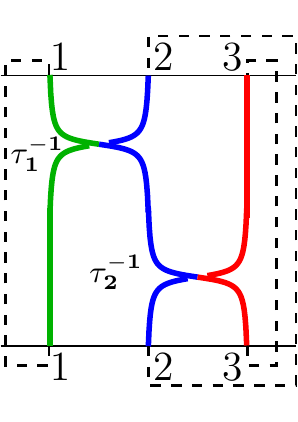}
(c)\includegraphics[scale=0.75]{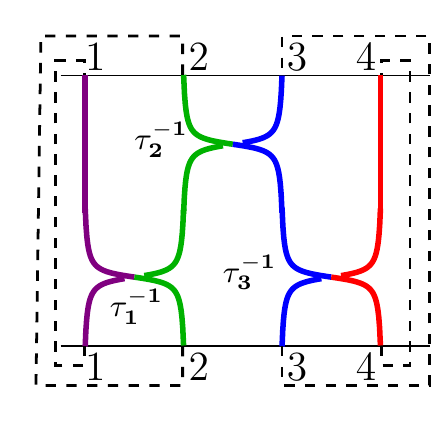}\\
(d)\includegraphics[scale=0.85]{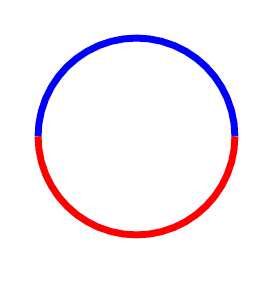}
(e)\includegraphics[scale=0.85]{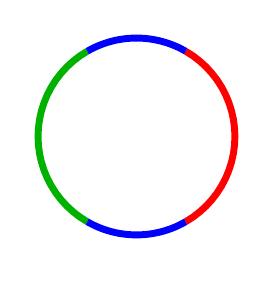}
(f)\includegraphics[scale=0.85]{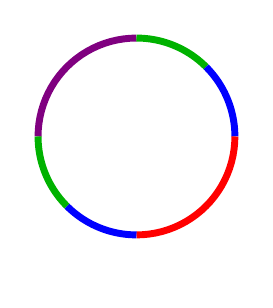}
\caption{The braid diagrams (with the braid closures by the dashed lines) obtained from the braiding of the complex energy bands of (a) nSSH2, (b) nSSH3,  and (c) nSSH4 models in (Re[$E_p(k)$], Im[$E_p(k)$], $k$) space. The corresponding knots called unknot for these cases are shown below. The small cut denotes that the $i^{th}$ band from left is going below the  $(i+1)^{th}$ band due to the braid operator $\tau^{-1}_i$. }
\label{fig:braid}
\end{figure}
 The braiding of multiple complex energy bands of the non-Hermitian SSH models in the composite metallic phases are shown in Figs.~\ref{fig:braid}(a,b,c).
 The corresponding braid words~\citep{hu_knots_2021,adams_knot_1994} to create these braids are $\tau_1^{-1}$, $\tau_2^{-1}\tau_1^{-1}$ and $\tau^{-1}_1\tau^{-1}_3\tau^{-1}_2$, respectively, for nSSH2, nSSH3 and nSSH4 model, where $\tau_i(\tau^{-1}_i)$ denotes the $i^{th}$ band from the left going above (below) the $(i+1)^{th}$ band. We note that the change in color of the bands/strands is due to the exchange of eigenmodes at the EPs~\citep{dembowski_encircling_2004,heiss_phases_1999}.
\begin{figure}
\vspace{15pt}
(a)\includegraphics[scale=0.575]{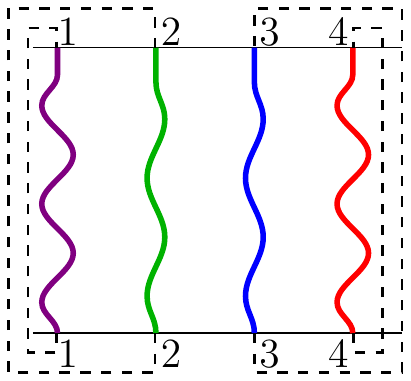}
(b)\includegraphics[scale=0.575]{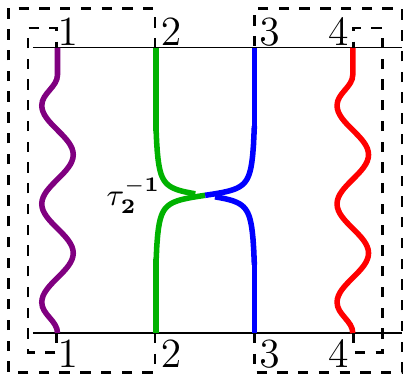}
(c)\includegraphics[scale=0.575]{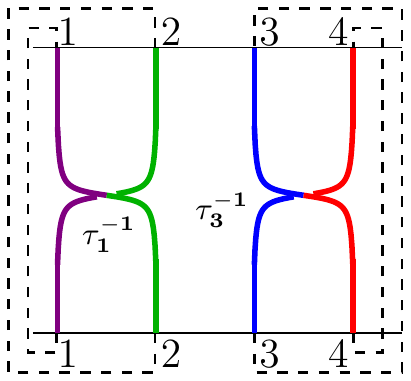}\\
\hspace{-5pt}(d)\includegraphics[scale=0.5]{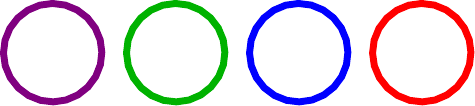}
(e)\includegraphics[scale=0.65]{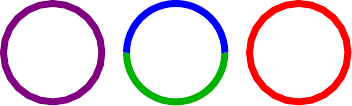}
\hspace{5pt}(f)\hspace{10pt}\includegraphics[scale=0.8]{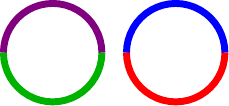}
\caption{The braid diagrams (with the braid closures by the dashed lines) obtained from the braiding of the complex energy bands of nSSH4 models in (Re[$E_p(k)$], Im[$E_p(k)$], $k$) space for (a) trivial and non-trivial topological insulators, (b) composite insulator $CI_z$ and (c) composite insulator $CI_m$. The corresponding knots (unlink and unknot) are shown below. The small cut denotes that the $i^{th}$ band from left is going below the  $(i+1)^{th}$ band due to the braid operator $\tau^{-1}_i$. }
\label{fig:braid1}
\end{figure}
Similar braids are also formed in the composite insulating phases of nSSH4 model by two participating bands shown in Figs.~\ref{fig:braid1}(b,c).  In the knot theory, disjoint circles of individual bands on the parametric plane form unlink, and the interlinked bands/braids form the different non-trivial knots. Therefore, all separable energy bands of the multipartite non-Hermitian SSH models in the trivial and non-trivial insulating phases (Fig.~\ref{fig:braid1}(a)) represent an unlink~\citep{wang_topological_2021,hu_knots_2021} shown in Fig.~\ref{fig:braid1}(d). Moreover, the composite metallic  phases with the inseparable energy bands have the simplest knot called unknot as shown in Figs.~\ref{fig:braid}(d,e,f), which resemble the respective composite loops in Figs.~\ref{fig:3},~\ref{fig:8},~\ref{fig:intro}(c). The composite loops in the composite insulating phases (Figs.~\ref{fig:intro}(d,b)) of nSSH4 model also represent the  unknot depicted in Figs.~\ref{fig:braid1}(e,f).

\bibliography{nssh4_prb1}

\end{document}